\documentclass[10pt,journal,compsoc]{IEEEtran}
\usepackage{amsmath,amsfonts}
\usepackage{algorithmic}
\usepackage{algorithm}
\usepackage{array}
\usepackage[caption=false,font=normalsize,labelfont=sf,textfont=sf]{subfig}
\usepackage{textcomp}
\usepackage{stfloats}
\usepackage{url}
\usepackage{verbatim}
\usepackage{graphicx}
\usepackage{cite}

\usepackage{booktabs} 

\usepackage{xcolor}

\newcommand{\rev}[1]{{{#1}}}

\newcommand{\techName}{\emph{StructLayoutFormer}}

\newcommand{\Align}{\textbf{Align}}
\newcommand{\Overlap}{\textbf{Overlap}}
\newcommand{\WLabel}{\textbf{W Label}}
\newcommand{\WBox}{\textbf{W Box}}

\newcommand{\SAlign}{\textbf{S-Align}}
\newcommand{\SOverlap}{\textbf{S-Overlap}}
\newcommand{\SInclusion}{\textbf{S-Inclusion}}
\newcommand{\WSLabel}{\textbf{W S-Label}}
\newcommand{\WSBox}{\textbf{W S-Box}}

\newcommand{\GenT}{\textbf{GenT}}
\newcommand{\GenTS}{\textbf{GenTS}}
\newcommand{\UGen}{\textbf{UGen}}
\newcommand{\Completion}{\textbf{Completion}}
\newcommand{\GenO}{\textbf{GenO}}
\newcommand{\StructTran}{\textbf{StructTran}}
\newcommand{\StructExtr}{\textbf{StructExtr}}

\newcommand{\EleMetric}{\textbf{element metrics}}
\newcommand{\StructMetric}{\textbf{structure metrics}}

\newcommand{\EleMetricCapital}{\textbf{Element metrics}}
\newcommand{\StructMetricCapital}{\textbf{Structure metrics}}
\newcommand{\NA}{-}

\usepackage{xspace}
\makeatletter
\DeclareRobustCommand\onedot{\futurelet\@let@token\@onedot}
\def\@onedot{\ifx\@let@token.\else.\null\fi\xspace}

\def\eg{{e.g}\onedot} 
\def\ie{{i.e}\onedot} 
 
\def\etc{{etc}\onedot} 
 
\def\etal{{et al}\onedot}
\makeatother

\usepackage{multirow}

\begin{document}

\title{\techName: \\ Conditional Structured Layout Generation via Structure Serialization and Disentanglement}

\author{Xin Hu, Pengfei Xu, Jin Zhou, Hongbo Fu, and Hui Huang
\IEEEcompsocitemizethanks{
\IEEEcompsocthanksitem Corresponding author: Pengfei Xu.

\IEEEcompsocthanksitem 
X. Hu (qzlyhx@hotmail.com), P. Xu (xupengfei.cg@gmail.com), J. Zhou (doudin2618@gmail.com), and H. Huang (hhzhiyan@gmail.com) are with the College of Computer Science and Software Engineering, Shenzhen University.
\IEEEcompsocthanksitem  H. Fu (fuplus@gmail.com) is with the Hong Kong University of Science and Technology. 
}
}


\markboth{Submitted to IEEE Transactions on Visualization and Computer Graphics}%
{}


\IEEEtitleabstractindextext{%
\begin{abstract}
Structured layouts are preferable in many 2D visual contents (\eg, GUIs, webpages) since the structural information allows convenient layout editing. 
Computational frameworks can help create structured layouts but require heavy labor input. Existing data-driven approaches are effective in automatically generating fixed layouts but fail to produce layout structures. 
We present \techName, a novel Transformer-based approach for conditional structured layout generation. We use a structure serialization scheme to represent structured layouts as sequences. To better control the structures of generated layouts, we disentangle the structural information from the element placements. 
Our approach is the first data-driven approach that achieves conditional structured layout generation and produces realistic layout structures explicitly.
We compare our approach with existing data-driven layout generation approaches by including post-processing for structure extraction.
Extensive experiments have shown that our approach exceeds these baselines in conditional structured layout generation. We also demonstrate that our approach is effective in extracting and transferring layout structures. 
The code is publicly available at 
{https://github.com/Teagrus/StructLayoutFormer}.

\end{abstract}

}

\maketitle



 
 
 
 


 
 
 



 




\section{Introduction}
\IEEEPARstart{G}{raphic}
layout plays an important role in graphic design. Traditionally, when designing 2D visual contents, \eg, posters, webpages, or GUIs, their layouts are created manually with interactive tools~\cite{xu2019global, zeidler2013auckland} or semi-automatically with computational frameworks~\cite{dayama2020grids, swearngin2020scout, xu2022hierarchical}. With the rise of learning techniques, this layout generation task can be achieved automatically in a data-driven manner. Existing data-driven approaches~\cite{arroyo2021variational, chai2023layoutdm, gupta2021layouttransformer, hui2023unifying, inoue2023layoutdm, jiang2022coarse, jyothi2019layoutvae, kikuchi2021constrained, kong2022blt, zheng2019content} have adopted GAN~\cite{goodfellow2014generative}, VAE~\cite{kingma2013auto}, GNN~\cite{scarselli2008graph}, Transformer~\cite{vaswani2017attention}, Diffusion model~\cite{ho2020denoising}, \etc, for the automatic layout generation task and obtained remarkable progress. 

\begin{figure}[t]
	\centering
	\includegraphics[width=\linewidth]{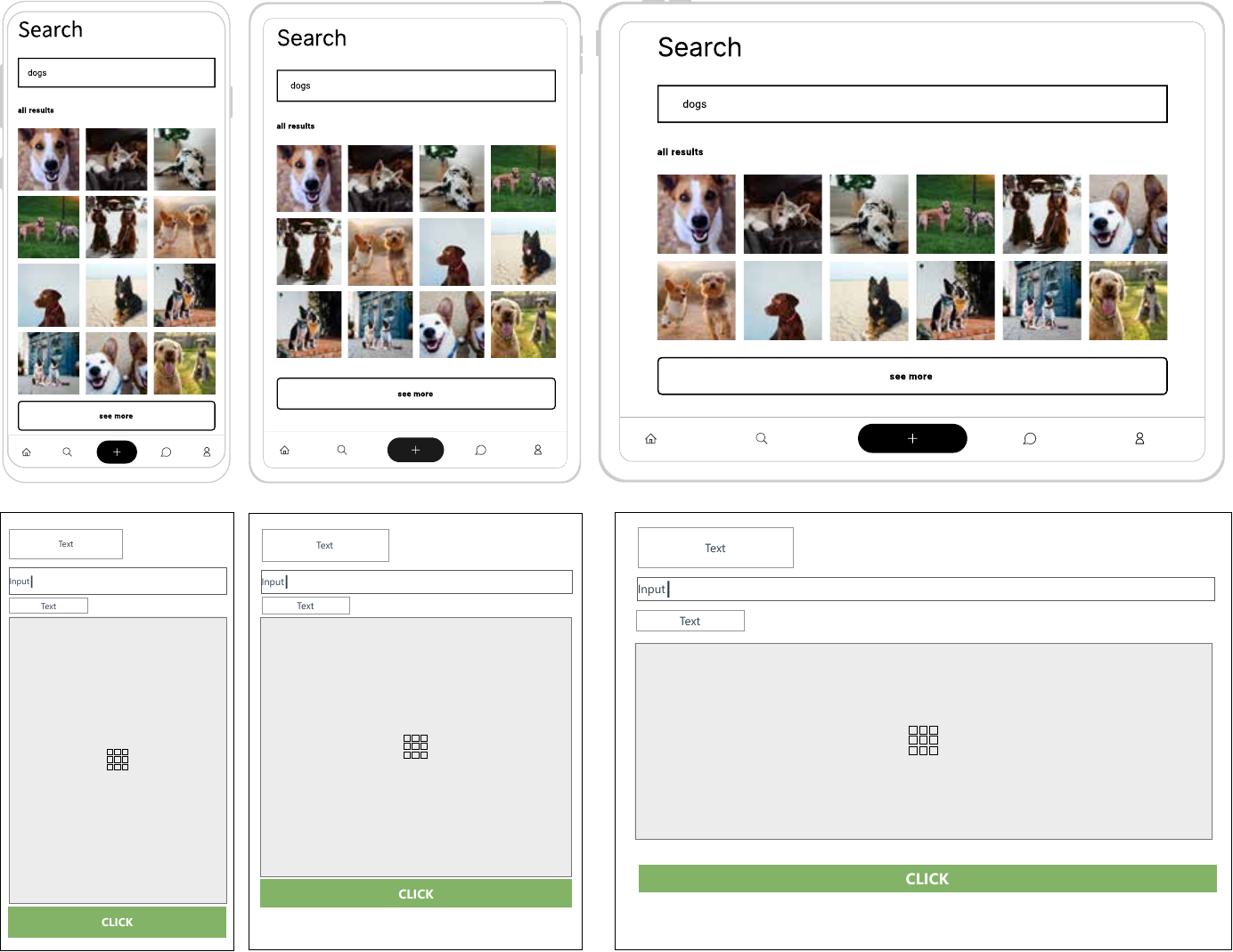}
	\caption{
        GUI layouts (top) and their major internal structures (below). In these examples, all images are under grid nodes, and their arrangements can be automatically adjusted to different screen sizes.
    }
	\label{fig:GUI}
\end{figure}

Compared with the computational methods for semi-automatic layout creation, data-driven approaches are more efficient in generating large numbers of layouts. However, existing data-driven approaches focus on producing fixed layouts represented as sets of bounding boxes, which is not suitable for further layout adjustment.
In contrast, computational methods can help create structured layouts containing relations among elements. This structural information enables structure-preserving manipulation of layouts. For example, the GUI layouts created with existing computational frameworks~\cite{zeidler2013auckland} contain relations among GUI elements. With such structural information, these GUI layouts can automatically adapt to different screen sizes~\cite{jiang2019orc} without manual input (see {Figure~\ref{fig:GUI}}). 
Nevertheless, these computational methods often require heavy user labor input. It would be meaningful to exploit learning techniques to generate structured layouts automatically.

\IEEEpubidadjcol
There exist several structured layout datasets, \eg, RICO~\cite{deka2017rico} and WebForest~\cite{kikuchi2021modeling}, in which layouts are represented as trees. A node of a layout tree represents a graphic element, which could be visible, \eg, a text box, or invisible, \eg, a linear arrangement. An internal node acts as a container of its children nodes.
For such structured data, recursive neural networks (RvNNs)~\cite{socher2011parsing} are often adopted for the automatic generation task. They have been successfully used for generating structured 3D shapes~\cite{li2017grass, mo2019structurenet, zhu2018scores} and indoor scenes~\cite{li2019grains}. 
However, RvNNs have the following limitations that prevent them from applying to the structured layout generation problem. 
First, RvNNs are inefficient in training since they require a bottom-up training procedure, which is not parallel for each sample. This is critical since a layout often contains many elements and has a complicated structure compared with structured 3D shapes or indoor scenes. 
Second, RvNNs prohibit messages from passing between different branches of a tree structure, restricting their learning capacity. 
Third, the existing RvNN-based methods are not suitable for conditional generation, thus limiting the application scenarios.
Compared with RvNN, Transformer is more potent in learning the arrangement patterns of graphic elements and can easily achieve conditional layout generation, as confirmed by existing approaches~\cite{gupta2021layouttransformer, kong2022blt, jiang2023layoutformer++, arroyo2021variational}. However, exploiting a Transformer architecture for the structured layout generation problem is nontrivial.

We propose \techName, a novel Transformer-based approach for conditional structured layout generation. In this work, a structured layout is represented as a layout tree. This representation has been adopted by existing layout datasets~\cite{deka2017rico, kikuchi2021modeling} and computational methods~\cite{dixon2011content, jiang2021reverseorc, jiang2022coarse, kikuchi2021modeling, o2014learning, xu2022hierarchical}. To adapt the Transformer architecture, we use a structure serialization scheme to map a layout tree to a sequence of tokens (see Figure~\ref{fig:serial}). This sequence contains all the structural information and can faithfully recover the layout tree. Our model produces such sequences autoregressively as the generated structured layouts. 
To better control the structure of the generated layouts, we disentangle the structural information of a layout from its element placement by constructing a latent space that embeds high-level layout structures. With this disentanglement, we can use a structure code as a condition for layout generation. It also enables our approach to support existing conditional layout generation tasks without considering a structured representation as input. Structure codes also bring more generation variety; our approach can produce different results under the same input conditions. Without structure codes, generating diverse structures can only be achieved through probabilistic sampling, which is less controllable.

We have extensively tested our approach on two structured layout datasets, \ie, RICO~\cite{deka2017rico} and 
\rev{WebForest}~\cite{kikuchi2021modeling}. The experiments include several conditional layout generation tasks. To better examine the effectiveness of our approach in structured layout generation, we compare our approach with the state-of-the-art layout generation approaches, including LayoutFormer++~\cite{jiang2023layoutformer++}, BLT~\cite{kong2022blt}, LayoutDM~\cite{inoue2023layoutdm}, and LayoutTransformer~\cite{gupta2021layouttransformer}. Different from our approach, these approaches cannot produce layout structures explicitly. However, they can find the arrangement patterns existing in the datasets. Therefore, we treat the internal nodes of layout trees as additional elements and use these approaches to produce layout structures implicitly. 
\rev{We also compare our approach with GTLayout~\cite{xu2024gtlayout}, an RvNN-based method that can produce layout structures.}
To quantitatively compare these approaches, besides the frequently-used metrics for measuring the quality of the element arrangement, we introduce additional metrics that measure the quality of the produced layout structures. The experiments show that our approach outperforms these approaches in the adopted metrics. We also demonstrate that our approach can extract and transfer layout structures.

\section{Related work}

\textbf{Structured layout creation.}
Many approaches have been proposed for structured layout creation due to the wide application of such layouts. Xu \etal~\cite{xu2019global} proposed a framework to create structured layouts interactively. The Auckland layout editor~\cite{zeidler2013auckland} was another interactive framework for structured GUI layout creation. Some computational approaches reduced the labor input for layout creation. O'Donovan \etal~\cite{o2014learning} presented an optimization method for generating structured grid layouts based on design principles. Kikuchi \etal~\cite{kikuchi2021modeling} proposed a method to transfer structures between existing webpage layouts. Girds~\cite{dayama2020grids} was a computational framework for structured GUI layout generation from heuristic rules. Scout~\cite{swearngin2020scout} was a system that helped designers explore structured GUI layouts. Xu \etal~\cite{xu2022hierarchical} proposed a method to create novel structured layouts via layout blending. Although these works help create structured layouts, they more or less require manual input or specification from people. In contrast, our approach is purely data-driven and may not require human intervention. 

\textbf{Learning-based layout generation.}
With the rise of learning techniques, many data-driven approaches have been proposed for layout generation. 
LayoutGAN~\cite{li2019layoutgan} was an early work exploiting GAN~\cite{goodfellow2014generative} for layout generation. LayoutGAN++~\cite{kikuchi2021constrained} also adopted GAN and improved the quality of generated layouts. LayoutVAE~\cite{jyothi2019layoutvae} adopted two VAEs~\cite{kingma2013auto} to generate layouts. LayoutTransformer~\cite{gupta2021layouttransformer} and VTN~\cite{arroyo2021variational} exploited Transformer~\cite{vaswani2017attention} for layout generation. Jiang \etal~\cite{jiang2022coarse} also exploited Transformer for layout generation. They realized the importance of layout structures. NDN~\cite{lee2020neural} adopted a GNN~\cite{scarselli2008graph} for layout generation. Recently, more works have focused on conditional layout generation. BLT~\cite{kong2022blt} extended BERT~\cite{kenton2019bert} for conditional layout generation. LayoutFormer++~\cite{jiang2023layoutformer++} could use geometric relations among elements as conditions for layout generation. The diffusion model~\cite{ho2020denoising} was also extended to the layout generation task, \eg, LayoutDMs~\cite{chai2023layoutdm, inoue2023layoutdm} and LDGM~\cite{hui2023unifying} achieving controllable layout generation with Diffusion models. 
%
These recent works have obtained remarkable progress in layout generation. However, none of them can produce layout structures explicitly. READ~\cite{patil2020read} exploited a binary tree structure for layout generation but could not be extended to general structure generation. GTLayout~\cite{xu2024gtlayout} adopted RvNN to achieve structured layout generation. However, it was designed for specific layout structures, including three types of element arrangements, making it difficult to adapt to real application scenarios that involve diverse structures. 
Jiang \etal~\cite{jiang2022coarse} proposed a VAE-based method for generating a two-layer structured layout.
In contrast, our data-driven approach explicitly achieves conditional generation of layouts with realistic structures.  

\rev{
Large language models (LLMs) have been extensively applied to layout generation tasks in recent studies. These approaches typically represent layouts in textual formats and design appropriate prompt instructions to leverage the reasoning capabilities of LLMs for generating layout samples. Most existing studies have not considered structured layout generation tasks~\cite{ feng2023layoutgpt, tang2023layoutnuwa, yang2024posterllava, yang2024llplace}. LayoutGPT~\cite{ feng2023layoutgpt} converts text inputs to image and indoor scene layouts. LayoutNUWA~\cite{ tang2023layoutnuwa} uses the SVG format to represent layouts and set unknown values as a mask token to achieve conditional layout generation. PosterLLaVa~\cite{ yang2024posterllava} combines background image and textual requirements to generate satisfied layouts.  LLplace~\cite{yang2024llplace} focuses on the 3D indoor scene layout generation and editing.  Lin \etal~\cite{lin2023parse} introduced Parse-Then-Place, a text-to-layout method that takes a rough description of the target layout, including the target's structure and contents, as input to generate a structured layout. This text-to-layout method cannot take precise constraints, \eg, element geometry, as conditions.
LLMs also demonstrate the capability of converting a webpage screenshot into a structured webpage represented as HTML code~\cite{laurenccon2024unlocking}. 
However, no LLM-based methods are currently capable of generating general structured layouts. The conversion of general structured layouts into textual representations for LLMs remains unexplored. The intermediate representation proposed in Parse-Then-Place~\cite{lin2023parse} can describe structured layouts in text. However, This representation is primarily employed to extract information from input text and lacks details to represent a complete layout. An additional Transformer decoder is required to transform the intermediate representation into specific layouts.
}

\textbf{Learning-based structure generation.}
Several works have achieved structured data generation. Most of them adopted RvNN~\cite{socher2011parsing} for their tasks. Li \etal~\cite{li2017grass} presented GRASS for structured 3D shape synthesis. Zhu \etal~\cite{zhu2018scores} presented SCORES for structured 3D shape composition. In these two works, shape structures are represented as binary trees. StructureNet~\cite{mo2019structurenet} exploited general trees for structured 3D shape generation. Li \etal~\cite{li2019grains} presented GRAINS, an RvNN-based method for indoor scene synthesis. 
As discussed earlier, RvNN has several limitations and cannot be applied to conditional structured layout generation. Li \etal~\cite{li2020auto} proposed a Transformer-based tree decoder on user interface completion task but did not achieve conditional generation.
Compared with these methods, our approach uses a Transformer architecture and succeeds in conditional structured layout generation.

\section{Approach}

Our approach adopts a Transformer architecture for structured layout generation. To adapt the Transformer architecture to our task, we use a structure serialization scheme to represent structured layouts as sequences. We also introduce a latent space to embed high-level layout structures. This latent space helps disentangle the structural information of layouts to achieve structure-conditioned layout generation. In the following, we first reiterate the layout representation adopted in our approach and introduce our serialization scheme that maps a layout tree to a sequence (Section~\ref{sec:lay_rep_serial}). We then explain our model architecture that helps achieve conditional structured layout generation (Section~\ref{sec:model}). Finally, we describe the training objective and training details of our model (Section~\ref{sec:training}).

\subsection{Layout representation and serialization}\label{sec:lay_rep_serial}

\begin{figure}
    \centering
    \includegraphics[width=\linewidth]{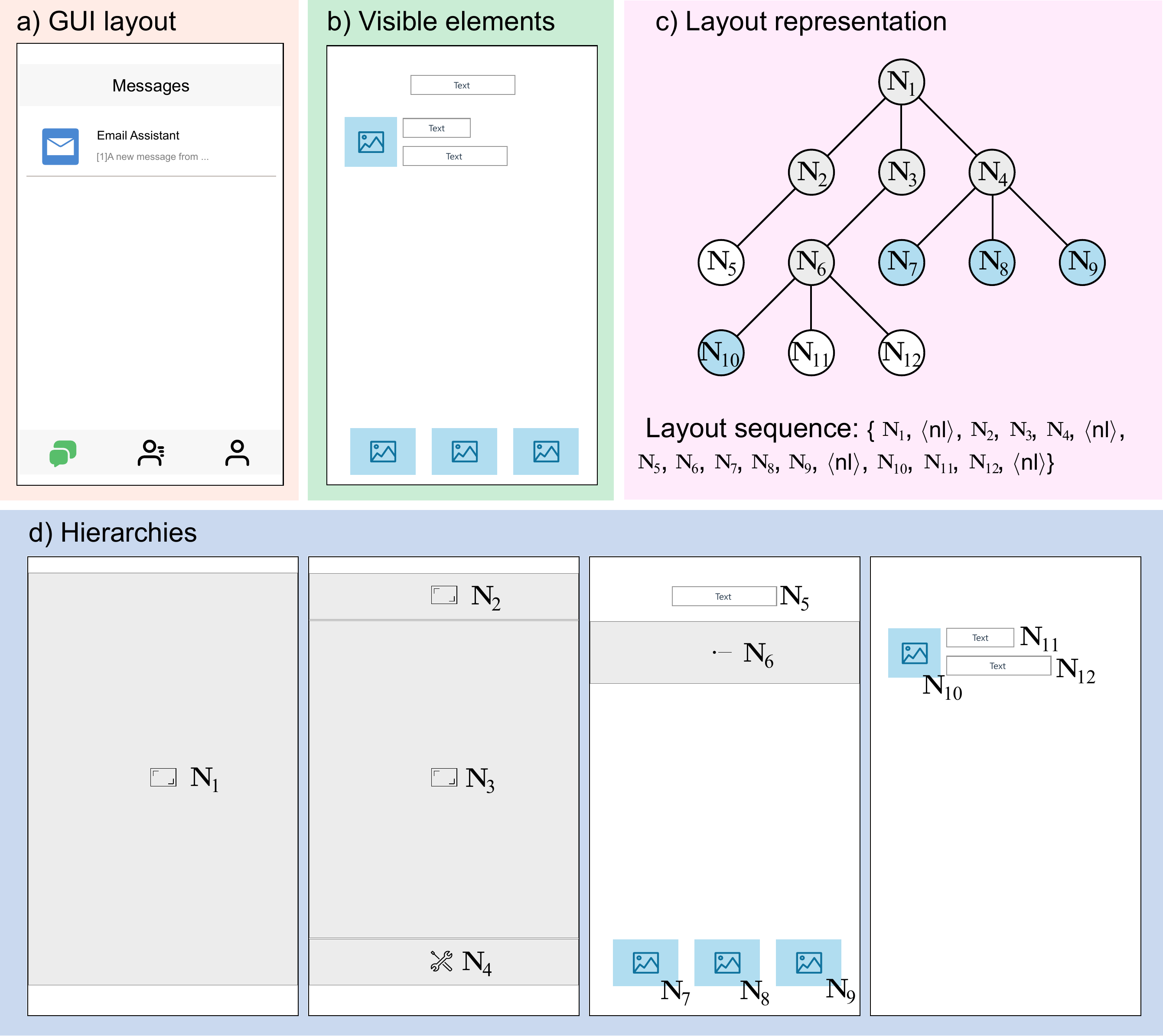}
    \caption{
    An illustration of layout representation and serialization. a) An example GUI layout. b) The layout's visible elements. c) The layout structure and the corresponding layout sequence. d) The visualization of the layout hierarchies.
    }
    
    \label{fig:serial}
\end{figure}

\begin{figure}
    \centering
    \includegraphics[width=\linewidth]{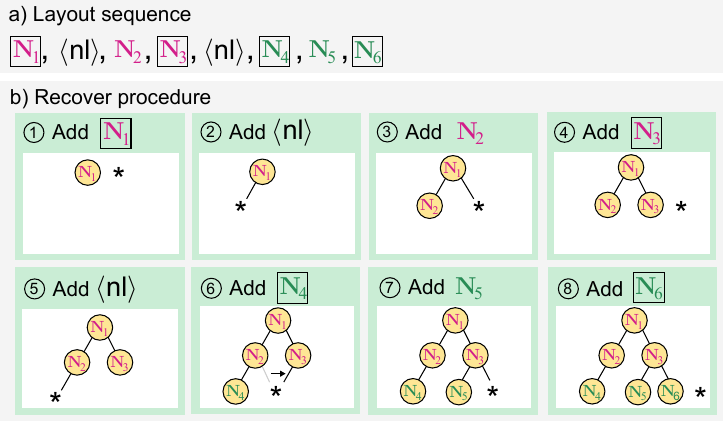}
    \caption{
    An example of recovering a layout sequence to a layout tree. a) In the layout sequence,  $\mathbf{N}_1$, $\mathbf{N}_2$, and $\mathbf{N}_3$ (in magenta) are internal nodes; $\mathbf{N}_4$, $\mathbf{N}_5$, and $\mathbf{N}_6$ (in green) are leaf nodes; $\mathbf{N}_1$, $\mathbf{N}_3$, $\mathbf{N}_4$, and $\mathbf{N}_6$ (with frames) are the last children of their parents. All this information is stored in the nodes. b) Detailed recovery procedure. The * symbol represents the position where the next predicted element will be placed.
    }
    \label{fig:se2tree}
\end{figure}

\textbf{Representation.}
Most existing data-driven approaches~\cite{arroyo2021variational, chai2023layoutdm, gupta2021layouttransformer, hui2023unifying, inoue2023layoutdm, jiang2022coarse, jyothi2019layoutvae, kikuchi2021constrained, kong2022blt} consider a layout as a set of bounding boxes. This representation is sufficient for unstructured layouts. However, for structured layouts, a more appropriate representation is the layout tree, which has been adopted by existing structured layout datasets~\cite{deka2017rico, kikuchi2021modeling} and computational frameworks~\cite{dixon2011content, jiang2021reverseorc, jiang2022coarse, kikuchi2021modeling, o2014learning, xu2022hierarchical}. Our approach adopts this representation for conditional structured layout generation. Specifically, a structured layout is represented as $\mathcal{T}=\{\mathbf{N}_i\}$, where $\mathcal{T}$ denotes a layout tree and $\mathbf{N}_i$ is a node of this layout tree. $\mathbf{N}_i$ contains the geometric and structural information of this node: $\mathbf{N}_i=[x_i, y_i, w_i, h_i, t_i, \{\mathbf{N}_j\}_i]$. $x_i$ and $y_i$ are the left and top coordinates of the node's bounding box. 
$w_i$ and $h_i$ are the bounding box's width and height. Following the previous works~\cite{gupta2021layouttransformer, kong2022blt, arroyo2021variational, jiang2023layoutformer++}, these four attributes are quantized. $t_i$ indicates the node's type, which can be a leaf node's semantic label or an internal node's organization type; $\{\mathbf{N}_j\}_i$ is a set containing this node's children. It is empty if this node is a leaf.   

\textbf{Serialization.}
The layout tree representation cannot be directly used in a Transformer architecture. We adopt the following serialization scheme to represent a tree structure with a sequence. As illustrated in Figure~\ref{fig:serial}, given a layout tree $\mathcal{T}=\{\mathbf{N}_i\}$, we first compose several sub-sequences, each of which consists of the nodes at the same level of the layout tree. 
In these sub-sequences, a node $\mathbf{N}_i$ does not contain its children set $\{\mathbf{N}_j\}_i$. This avoids the recursive representation problem. To retain the structural information, we add two binary variables $b_i^{1}$ and $b_i^{2}$, {and then $\mathbf{N}_i=[x_i, y_i, w_i, h_i, t_i, b_i^{1}, b_i^{2}]$.} 
$b_i^{1}$ indicates whether $\mathbf{N}_i$ is a leaf node or an internal node. $b_i^{2}$ indicates whether $\mathbf{N}_i$ is the last child node of its parent. With these two binary variables, we can recover the relations among the nodes in different sub-sequences. To represent the structured layout completely, we concatenate these sub-sequences in the order of their levels in the layout tree. To retain the level information, we add an extra token $\langle \mathrm{nl} \rangle$ between the adjacent sub-sequences. Figure \ref{fig:se2tree} shows an example of recovering a layout sequence back to a layout tree.

\begin{figure*}
    \centering
    \includegraphics[width=\textwidth]{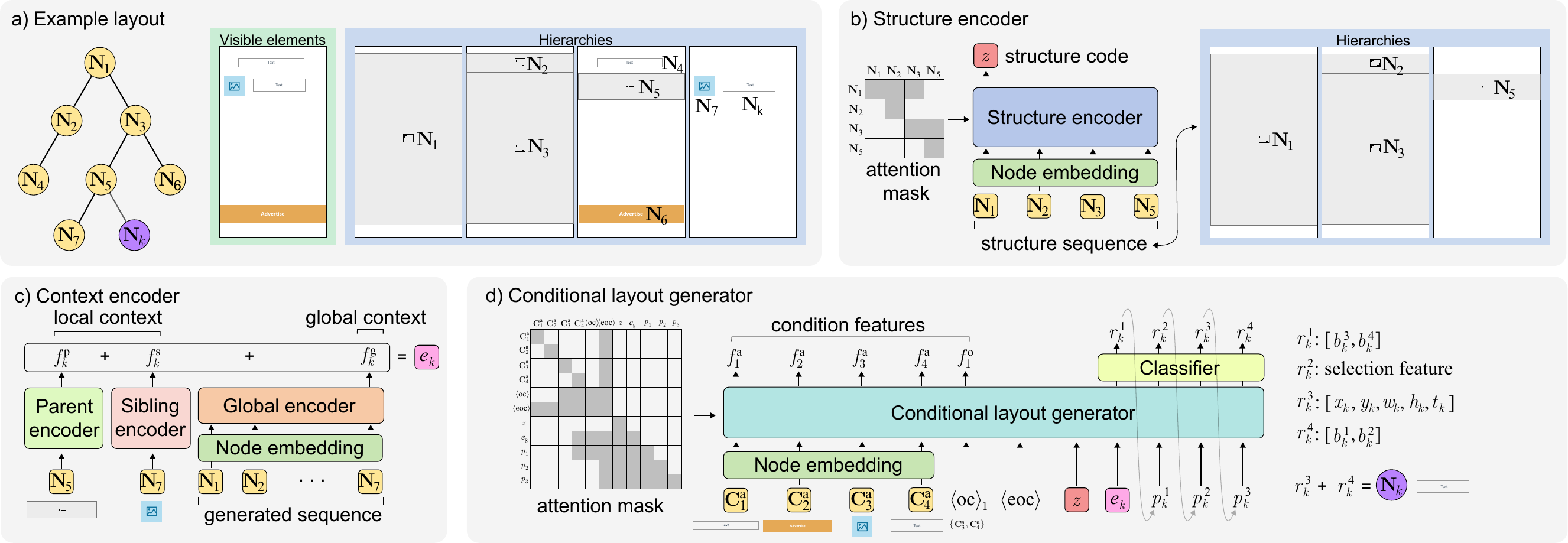}
    \caption{
     An overview of our model. a) An example of layout generation conditioned on element types and sizes. The purple node is the target of the current round of node prediction. b) Structure encoder is a VAE that encodes a structure sequence into a latent structure code $z$. c) Context encoder encodes the context information of $\mathbf{N}_k$(the node model predicts now) to a context code $e_k$. The context information includes its parent, sibling, and the predicted nodes. d) Conditional layout generator selects the proper condition according to structure $z$ and the context code $e_k$ then predicts attributes of the new element $\mathbf{N}_k$.
    }
    \label{fig:model}
\end{figure*}

\subsection{Model architecture}\label{sec:model}
Figure~\ref{fig:model} illustrates our model architecture. Our model produces a layout sequence autoregressively to generate a structured layout. 
It has three components: a conditional layout generator, a structure encoder, and a context encoder.
The input of the conditional layout generator consists of three parts. The first part is an element condition sequence. It contains the constraints on the elements of the generated layout, \eg, the element types or sizes. The second part is a structure code. It determines the structure of the generated layout. During training, this code is obtained by encoding the high-level structure of a sample with the structure encoder. The third part is a context code. This code changes dynamically in the autoregressive generation procedure. For each generation step, this code is updated by feeding the already generated layout sequence to the context encoder. In the following, we describe our model in detail.

\textbf{Element condition.}
The element condition contains the constraints on the elements of the generated layout. Our approach supports two types of conditions: attribute conditions and organization conditions.  An attribute condition specifies an element's attributes, \ie, position, size, and type. It is defined as
\rev{$\mathbf{C}^{\mathrm{a}}  = [x, y, w, h, t]$}.
Note that an attribute condition may only specify a subset of an element's attributes. We thus introduce a mask token $\langle \mathrm{m} \rangle$ to replace the unspecified attributes. For each attribute condition, we convert it into a token via an embedding network, which will be described later. Then, a list of attribute conditions becomes a list of tokens. 

An organization condition constrains a set of nodes to be siblings in the generated structured layout. It is difficult to embed such a constraint in a token. 
%
Instead, we use a fixed token $\langle \mathrm{oc} \rangle$ to indicate an organization condition and achieve its constraint via an attention mask. This attention mask only allows the condition tokens constrained by this organization condition to pass messages to this token $\langle \mathrm{oc} \rangle$. We will explain the details later when introducing the conditional layout generator.  
Combining all the attribute and organization conditions results in an element condition sequence.
We append an extra $\langle \mathrm{eoc} \rangle$ token to indicate the end of conditions.

\textbf{Structure encoder.}
We introduce the structure encoder to disentangle the structural information of a layout from its element placement. For a structured layout represented as a layout tree, its internal nodes indicate how the graphic elements are hierarchically organized, and its leaf nodes reflect the element placement. 
In addition, a higher-level internal node, \ie, the one closer to the root, contributes more to the structure, and a lower-level internal node influences more on the element placement.
We thus extract the structural information of the layout from its internal nodes. Specifically, we remove the leaf nodes in the layout sequence to obtain a structure sequence. This structure sequence is then fed to the structure encoder to get a structure code. The structure encoder adopts a Transformer-VAE architecture, similar to VTN~\cite{arroyo2021variational}. To better capture the structural information, we introduce an attention mask in this structure encoder. This attention mask only allows the message of a node to pass to its parent node and itself, implicitly enhancing the influence of the high-level nodes.

\textbf{Context encoder.}
Our model produces a layout sequence autoregressively. Each time when generating a new node $\mathbf{N}_k$ of the layout sequence, the previously generated partial sequence is fed to the context encoder to obtain a context code. Since the generated sequence contains the structural information (see Section~\ref{sec:lay_rep_serial}), we can determine the relation between $\mathbf{N}_k$ and the previously generated nodes. Based on this structural information, we define local and global contexts. The local context is estimated by feeding the parent node and the most recently generated sibling node of $\mathbf{N}_k$ to two separate FC layers and then adding the extracted features $f^\mathrm{p}_k$ and $f^\mathrm{s}_k$. The global context $f^\mathrm{g}_k$ is estimated by feeding the generated nodes to a Transformer block. We then add these two context features to obtain the final context code.

\textbf{Conditional layout generator.}
The conditional layout generator is a Transformer block. In each autoregressive generation step, it consumes an element condition sequence, a structure code, and a context code to generate a new node $\mathbf{N}_k$. During the generation, the element conditions will be satisfied gradually. As the conditions change, we do not modify the element condition sequence. Instead, we apply an attention mask to indicate the updated conditions. For example, if a condition is already satisfied, the message of the corresponding token cannot pass to other tokens. This attention mask also provides other restrictions. We describe this attention mask as follows: (a) an element condition token only accepts messages from itself and the $\langle \mathrm{eoc} \rangle$ token; (b) an organization condition token accepts messages from itself, the element tokens constrained by this organization condition, and the $\langle \mathrm{eoc} \rangle$ token; (c) the $\langle \mathrm{eoc} \rangle$ token accepts messages from all tokens in the element condition sequence; (d) the structure code token only accepts messages from itself; (e) the context code token accepts messages from unsatisfied condition tokens, the $\langle \mathrm{eoc} \rangle$ token, the structure code token, and itself.


Generating a new node $\mathbf{N}_k$ is achieved in four sub-steps whose outputs are $r^1_k$, $r^2_k$, $r^3_k$, and $r^4_k$ respectively (Figure~\ref{fig:model}).  
The input sequences in these steps have a common part \{$\mathbf{C}_1^{\mathrm{a}},...,\mathbf{C}_n^{\mathrm{a}}, \mathrm{\langle oc\rangle}_1,...,\mathrm{\langle oc \rangle}_m,\mathrm{\langle eoc \rangle},z,e_k$\} which contains $n$ attribute conditions, $m$ organization conditions, an $\langle \mathrm{eoc} \rangle$ mark, a structure code $z$, and a context code $e_k$. We use $\mathbf{S}_{\mathrm{c}}$ to represent this sequence in the following descriptions. In \emph{Substep 1}, the conditional layout generator takes  $\mathbf{S}_{\mathrm{c}}$ as input to generate a feature $r^1_k$, which is
then converted into two binaries $b^3_k$ and $b^4_k$ through an MLP classifier. 
The first binary indicates whether $\mathbf{N}_k$ is an $\langle \mathrm{nl} \rangle$, and the second indicates whether $\mathbf{N}_k$ should satisfy a condition. 
If $\mathbf{N}_k$ is an $\langle \mathrm{nl} \rangle$, then this round of generation stops. Otherwise, we check if $\mathbf{N}_k$ should satisfy a condition. If yes, we continue \emph{Substep 2}; if no, we go to \emph{Substep 3}. 

In \emph{Substep 2}, we convert $r^1_k$ into a token $p^1_k$ with an MLP 
and append it to the input sequence to obtain \{$\mathbf{S}_{\mathrm{c}},p^1_k$\}. The generator produces a selection feature $r^2_k$ and calculates its distances to condition features including $f_1^\mathrm{a}...f_n^\mathrm{a}$ and $f_1^\mathrm{o}...f_m^\mathrm{o}$. \rev{The condition features are the outputs of the conditional layout generator at the corresponding positions of the input conditions.}
We then aggregate all distances to form a selection categorical probability distribution through a softmax function and finally sample a target condition feature. We force the predicted node in this step to match the selected condition.

In \emph{Substep 3}, 
we define a new token $p^2_k$ as the selected condition token or a zero token if no condition is selected. We then append it to the input sequence. Then the input sequence becomes \{$\mathbf{S}_{\mathrm{c}},p^1_k,p^2_k$\}.
The generator takes this sequence and produces a feature $r^3_k$. Finally we use feature extraction MLPs on $r^3_k$ to get $[x_k, y_k, w_k, h_k, t_k]$ of $\mathbf{N}_k$.


In \emph{Substep 4},
we convert $[x_k, y_k, w_k, h_k, t_k]$ into a new token $p^3_k$ and append $p^3_k$ to the input sequence to obtain 
\{$\mathbf{S}_{\mathrm{c}},p^1_k,p^2_k,p^3_k$\}. 
The generator consumes this sequence to produce a feature $r^4_k$.  We then extract two binaries $b^1_k$ and $b^2_k$. $b^1_k$ decides whether $\mathbf{N}_k$ is a leaf node and $b^2_k$ indicates whether $\mathbf{N}_k$ is the last child of its parent. 
After this substep, we finish the generation of the new node $\mathbf{N}_k=[x_k, y_k, w_k, h_k, t_k, b^1_k, b^2_k]$.


\textbf{Node embedding.}
In our model, we treat a node of a layout tree as one single token. Such a token is obtained by feeding a node 
\rev{$\mathbf{N}=[x, y, w, h, t, b^1,  b^2]$}
to an FC layer. This helps reduce the sequence lengths. This node embedding is adopted in the structure encoder and the context encoder. When preparing the element condition sequence, an attribution condition
\rev{$\mathbf{C}^{\mathrm{a}} = [x, y, w, h, t]$}
is converted into a token via this embedding layer.

\subsection{Training}\label{sec:training}

We train our model on structured layout datasets and adopt the teacher-forcing training technique. The structure encoder, the context encoder, and the conditional layout generator are trained together. Our generator produces a node $\mathbf{N}_k$ or a next-level token $\langle\mathrm{nl}\rangle$ in each generation step. Since the attributes of this node are all quantized, we adopt cross-entropy losses for training. In addition, since the structure encoder adopts a Transformer-VAE architecture, we include a KL-divergence loss.

\begin{table*}[t]
    \centering
    
    \begin{tabular}{ p{2cm} p{15cm}  }
    \toprule
    \textbf{Datasets} & \textbf{Element Types}\\
    \hline
    \parbox[c][1.3cm][c]{3cm}{RICO}& \parbox[c][1.3cm][c]{15cm}{View, LinearLayout, RelativeLayout, FrameLayout, ViewPager, ListView, GridView, Toolbar, Card, ListItem, Drawer, RecyclerView, WebView, Advertisement, TextButton, ButtonBar, Icon, DatePicker, Modal, Text, Image, Video, Checkbox, Input, BackgroundImage, NumberStepper, MapView, OnOffSwitch, Slider, RadioButton, PagerIndicator, MultiTab, BottomNavigation}\\
    \hline
    WebForest & Root, Container, Image, Text, Button, Graphic, Input \\
    \bottomrule
    
    \end{tabular}
    
    \vspace{1mm}
    \caption{The element types in two datasets.}
    \label{tab:eletype}
\end{table*}   

\section{Experiments}

\subsection{Setups}

\begin{figure}[t]
	\centering
	\includegraphics[width=\linewidth]{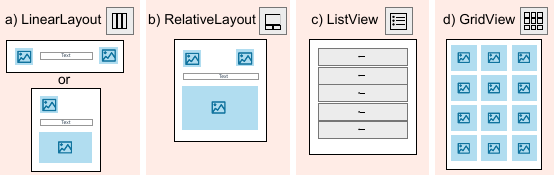}
	\caption{
        Examples of four typical internal nodes. a) LinearLayout. All elements are arranged linearly in the horizontal or vertical direction. b) RelativeLayout. Elements can be represented flexibly in a non-linear manner. c) ListView. Elements are arranged in a list and are usually of equal size. d) GridView. Elements are arranged in a grid and are usually of equal size.
    }
	\label{fig:inodeExample}
\end{figure}

\textbf{Datasets.}
We adopt two structured layout datasets in our experiments: RICO~\cite{deka2017rico} and WebForest~\cite{kikuchi2021modeling}. RICO contains more than 66K mobile GUI layouts with well-defined structures. This dataset has been adopted for the experiments in existing works~\cite{jiang2023layoutformer++, gupta2021layouttransformer, arroyo2021variational, kong2022blt, inoue2023layoutdm}. However, all these works focus on the element placement only and neglect the layout structures.
In contrast, we test our approach on this dataset for structured GUI layout generation. \rev{WebForest} contains 4.5K webpage layouts with structures. This dataset was proposed to evaluate a computational framework~\cite{kikuchi2021modeling} for creating structured layouts. We test our approach on this dataset for structured webpage layout generation. 
Table \ref{tab:eletype} shows the element types in these two datasets. Figure~\ref{fig:inodeExample} shows four typical internal nodes in RICO. Google's Android widget reference\footnote{developer.android.com/reference/android/widget/package-summary} gives a more detailed explanation for internal nodes. The training/testing ratio for both datasets is 9:1.

\textbf{Baselines.}
To the best of our knowledge, no existing data-driven approach achieves general conditional structured layout generation. READ~\cite{patil2020read} adopts RvNN for layout generation. However, it considers a layout structure as a binary tree and cannot generate general layout structures. GTLayout~\cite{xu2024gtlayout} also adopts RvNN and can produce layout structures. However, the produced structure is limited to three types of arrangements, \ie, vertical arrangement, horizontal arrangement, and stack arrangement.
\rev{Parse-Then-Place~\cite{lin2023parse} is another approach capable of generating structured layouts. It takes text descriptions, which contain explicit structural information, as input and has a distinct setting to our approach.}
On the other hand, the existing approaches can find the arrangement patterns in the datasets. For a structured layout represented as a layout tree, a parent node and its child node often follow certain arrangement patterns, \ie, this parent node often includes its child node. If we treat the internal nodes of layout trees as additional elements and assign them corresponding labels, the existing approaches may produce layout structures implicitly. 

Specifically, we achieve structured layout generation in three steps with these approaches. 
First, we focus on the leaves of layout trees and perform layout generation with these approaches. This is the same as the traditional layout generation task for unstructured layouts. Second, we use the leaves as the condition to predict the internal nodes of layout trees. Note that, in the first and second steps, we train two separate models for each approach. After obtaining the leaves and internal nodes, we determine the tree structure with the following procedure: for each element, its parent is defined as the one that has an internal label, has a larger size than this element, and covers this element most.

We thus compare our approach with the following state-of-the-art layout generation approaches, including LayoutFormer++~\cite{jiang2023layoutformer++}, BLT~\cite{kong2022blt}, LayoutDM~\cite{inoue2023layoutdm}, and LayoutTransformer~\cite{gupta2021layouttransformer}. 
We also tried VTN~\cite{arroyo2021variational}, but its training failed due to memory limit. 
GTLayout~\cite{xu2024gtlayout} can also produce structured layouts but can not adapt to general layout structures. We compare our approach with GTLayout in its structure setting in a separate experiment (Section~\ref{sec:comp-gtlayout}).

\begin{table*}[t]
    \centering
        \resizebox{\textwidth}{!}{
        \begin{tabular}{ l  l  r  r  c  c  c  r  r  r  r }
            \toprule
            & & \multicolumn{5}{ c } {\StructMetricCapital}  & \multicolumn{4}{ c } {\EleMetricCapital} \\
            \cmidrule(lr){3-7}
            \cmidrule(lr){8-11}
            
            Tasks & Approaches & 
            
            \WSLabel\ $\downarrow$ & \WSBox\ $\downarrow$& \SInclusion\ $\uparrow$& \SAlign\ $\downarrow$& \SOverlap\ $\downarrow$ & \Align\ $\downarrow$& \Overlap\ $\downarrow$& \WLabel\ $\downarrow$& \WBox\ $\downarrow$ \\
            \hline
            \multirow{4}{*}{\GenTS} 
            & Ours &
                \textbf{4.90} & \textbf{0.033} & \underline{0.927} & 0.0032 & \textbf{0.049} & 0.0019 & \textbf{0.024} & \underline{0.083} & \underline{0.026}
            \\
                
            & LayoutFormer++ &
                54.12 & 0.088 & 0.888 & \underline{0.0031} & \underline{0.067} & 0.0014 & \underline{0.032} & 0.272 & 0.053
            \\

            & BLT &
                16.70 & 0.139 & 0.318 & \textbf{0.0017} & 0.207 & \textbf{0.0007} & 0.114 & 0.504 & 0.121
            \\

            & LayoutDM &
                \underline{7.57} & \underline{0.078} & \textbf{0.939} & 0.0042 & 0.114 & \underline{0.0008} & 0.049 & \textbf{0.025} & \textbf{0.013}
            \\
            
            \hline
    
            \multirow{4}{*}{\GenT} & Ours &
                \textbf{4.63} & \textbf{0.038} & \underline{0.942} & 0.0023 & \underline{0.052} & 0.0034 & \textbf{0.022} & \underline{0.164} & \underline{0.040}
            \\

            & LayoutFormer++ &
                57.37 & 0.112 & 0.868 & \underline{0.0022} & \textbf{0.048} & \underline{0.0006} & \underline{0.017} & 0.351 & 0.071
            \\
            
            & BLT &
                13.87 & 0.198 & 0.358 & \textbf{0.0005} & 0.240 & \textbf{0.0004} & 0.141 & 0.402 & 0.232
            \\

            & LayoutDM &
                 \underline{6.01} & \underline{0.085} & \textbf{0.943} & 0.0042 & 0.117 & 0.0009 & 0.047 & \textbf{0.05} & \textbf{0.012}
            \\
            \hline
            \multirow{3}{*}{\UGen} & Ours &
                \underline{11.10} & \textbf{0.052} & \underline{0.908} & \textbf{0.0018} & \underline{0.057} & \textbf{0.0003} & \underline{0.019} & 0.764 & \underline{0.042}
            \\
            
            & LayoutFormer++&
                58.18 & 0.134 & 0.890 & \underline{0.0020} & \textbf{0.038} & \underline{0.0005} & \textbf{0.009} & \underline{0.234}& 0.079
            \\

            & LayoutDM &
               \textbf{6.77} & \underline{0.079} & \textbf{0.941} & 0.0049 & 0.123 & 0.0009 & 0.051 & \textbf{0.089} & \textbf{0.014}
            \\
            
            \hline
    
            \multirow{4}{*}{\Completion} & Ours&
                \textbf{3.62} & \textbf{0.053} & 0.863 &  \underline{0.0021} & \textbf{0.026} & 0.0070 & \textbf{0.006} & \underline{0.205} & \underline{0.023}
            \\
            
            & LayoutFormer++ &
                38.73 & 0.096 & \textbf{0.939} & 0.0066 & \underline{0.072} & 0.0028 & \underline{0.033} & 0.592 & 0.060
            \\
                
            & BLT &
                18.89 & 0.123 & 0.322 & \textbf{0.0016} & 0.213 & \underline{0.0008} & 0.112 & 0.561 & 0.135
            \\
            & LayoutDM &
                \underline{5.08} & \underline{0.094} & \underline{0.938} & 0.0040 & 0.115 & \textbf{0.0007} & 0.040 & \textbf{0.091} & \textbf{0.017}
            \\

            \hline

            \multirow{4}{*}{\StructExtr} & Ours&
                \textbf{2.35} & \textbf{0.041} & 0.801 & 0.0047 & \underline{0.050}
                & \NA & \NA & \NA & \NA
            \\
            
            & LayoutFormer++ &
                57.73 & 0.114 & \underline{0.836} & \underline{0.0038} & \textbf{0.046}
                & \NA & \NA & \NA & \NA
            \\
                
            & BLT &
                17.41 & 0.120 & 0.311 & \textbf{0.0030} & 0.167
                & \NA & \NA & \NA & \NA
            \\

            & LayoutDM &
                \underline{4.82} & \underline{0.084} & \textbf{0.942} & 0.0041 & 0.105
                & \NA & \NA & \NA & \NA
            \\

            \bottomrule
        \end{tabular}
    }
    \vspace{1mm}
    \caption{Quantitative comparisons of \GenT, \GenTS, \UGen, \Completion\ and \StructExtr\ on RICO.}
    \label{tab:rico}
\end{table*}    

\begin{table*}[t]
    \centering
    \resizebox{\textwidth}{!}{
        \begin{tabular}{ l  l  r  r  c  c  c  r  r  r  r }
            \toprule
            & & \multicolumn{5}{ c } {\StructMetricCapital}  & \multicolumn{4}{ c } {\EleMetricCapital} \\
            \cmidrule(lr){3-7}
            \cmidrule(lr){8-11}
            
            Tasks & Approaches & 
            \WSLabel\ $\downarrow$ & \WSBox\ $\downarrow$& \SInclusion\ $\uparrow$& \SAlign\ $\downarrow$& \SOverlap\ $\downarrow$ & \Align\ $\downarrow$& \Overlap\ $\downarrow$& \WLabel\ $\downarrow$& \WBox\ $\downarrow$\\
            \hline
            \multirow{4}{*}{\GenTS} & Ours &
                \textbf{0.84} & \textbf{0.056} & 0.847 & 0.0048 & \textbf{0.052} & 0.0022 & \textbf{0.029} & \textbf{0.018} & 0.037
            \\
                
            & LayoutFormer++ &
                \underline{1.09} & 0.095 & \textbf{0.989} & 0.0118 & \underline{0.067} & 0.0053 & \textbf{0.029} & 0.084 & \underline{0.027}
            \\

            & BLT &
                2.95 & 0.232 & 0.100 & \textbf{0.0020} & \underline{0.067} & \textbf{0.0006} & \underline{0.043} & 0.144 & 0.096
            \\

            & LayoutDM &
            7.57 & \underline{0.078} & \underline{0.939} & \underline{0.0042} & 0.114 & \underline{0.0008} & 0.049 & \underline{0.025} &\textbf{0.013}
            \\
            \hline
    
            \multirow{4}{*}{\GenT} & Ours &
                \textbf{0.92} & \textbf{0.047} & 0.862 & 0.0050 & \underline{0.048} & 0.0020 & 0.027 & \textbf{0.025} & 0.047
            \\
            
            & LayoutFormer++ &
                \underline{1.39} & 0.098 & \textbf{0.988} & 0.0092 & 0.058 & 0.0022 & \textbf{0.016} & 0.124 & \underline{0.023}
            \\

            & BLT &
                3.24 & 0.239 & 0.085 & \textbf{0.0009} & \textbf{0.038} & \textbf{0.0007} & \underline{0.026} & 0.093 & 0.13
            \\

            & LayoutDM &
            6.01 & \underline{0.085} & \underline{0.943} & \underline{0.0042} & 0.117 & \underline{0.0009} & 0.047 & \underline{0.053} & \textbf{0.012}
            \\
            \hline
            \multirow{3}{*}{\UGen} & Ours &
                \textbf{0.74} & \textbf{0.056} & 0.827 & \textbf{0.0043} & \underline{0.058} & \underline{0.0010} & \underline{0.017} & \underline{0.110} & 0.035
            \\
            
            & LayoutFormer++&
           \underline{1.04} & 0.108 & \textbf{0.993} & 0.0074 & \textbf{0.042} & 0.0018 & \textbf{0.005} & 0.141 & \underline{0.029}

            \\

            & LayoutDM &
            6.77 & \underline{0.079} & \underline{0.941} & \underline{0.0049} & 0.123 & \textbf{0.0009} & 0.051 & \textbf{0.089} & \textbf{0.014}
            \\
            
            \hline
    
            \multirow{4}{*}{\Completion} & Ours&
                \textbf{1.18} & \textbf{0.044} & 0.783 & 0.0106 & \textbf{0.012} & 0.0024 & \textbf{0.002} & \textbf{0.018} & \underline{0.027}
            \\
            
            & LayoutFormer++ &
                \underline{1.56} & 0.111 & \textbf{0.992} & 0.0105 & 0.046 & 0.0031 & \underline{0.010} & \underline{0.077}& 0.039
            \\
                
            & BLT &
                3.55 & 0.254 & 0.065 & \textbf{0.0014} & \underline{0.029} & \underline{0.0009} & 0.024 & 0.333 & 0.140
            \\

            & LayoutDM &
            5.08 & \underline{0.094} & \underline{0.938} & \underline{0.0040} & 0.115 & \textbf{0.0007} & 0.040 & 0.091 & \textbf{0.017}
            \\

            \hline
    
            \multirow{4}{*}{\StructExtr} & Ours&
                \underline{2.34}  & \underline{0.106} & 0.798 & 0.0105 & \textbf{0.007}
                & \NA & \NA & \NA & \NA
            \\
            
            & LayoutFormer++ &
                \textbf{1.14} & \textbf{0.092} & \textbf{0.991} & 0.0079 & 0.039 
                & \NA & \NA & \NA & \NA
            \\
                
            & BLT &
                2.59 & 0.247 & 0.083 & \textbf{0.0021} & \underline{0.029}
                & \NA & \NA & \NA & \NA
            \\

            & LayoutDM &
                2.65 & 0.145 & \underline{0.937} & \underline{0.0029} & 0.044
                & \NA & \NA & \NA & \NA
            \\

            \bottomrule
        \end{tabular}
    }
    \vspace{1mm}
    \caption{Quantitative comparisons of \GenT, \GenTS, \UGen, \Completion\ and \StructExtr\ on WebForest.}
    \label{tab:web}
\end{table*}

\begin{figure*}
    \centering
    \includegraphics[height=230mm]{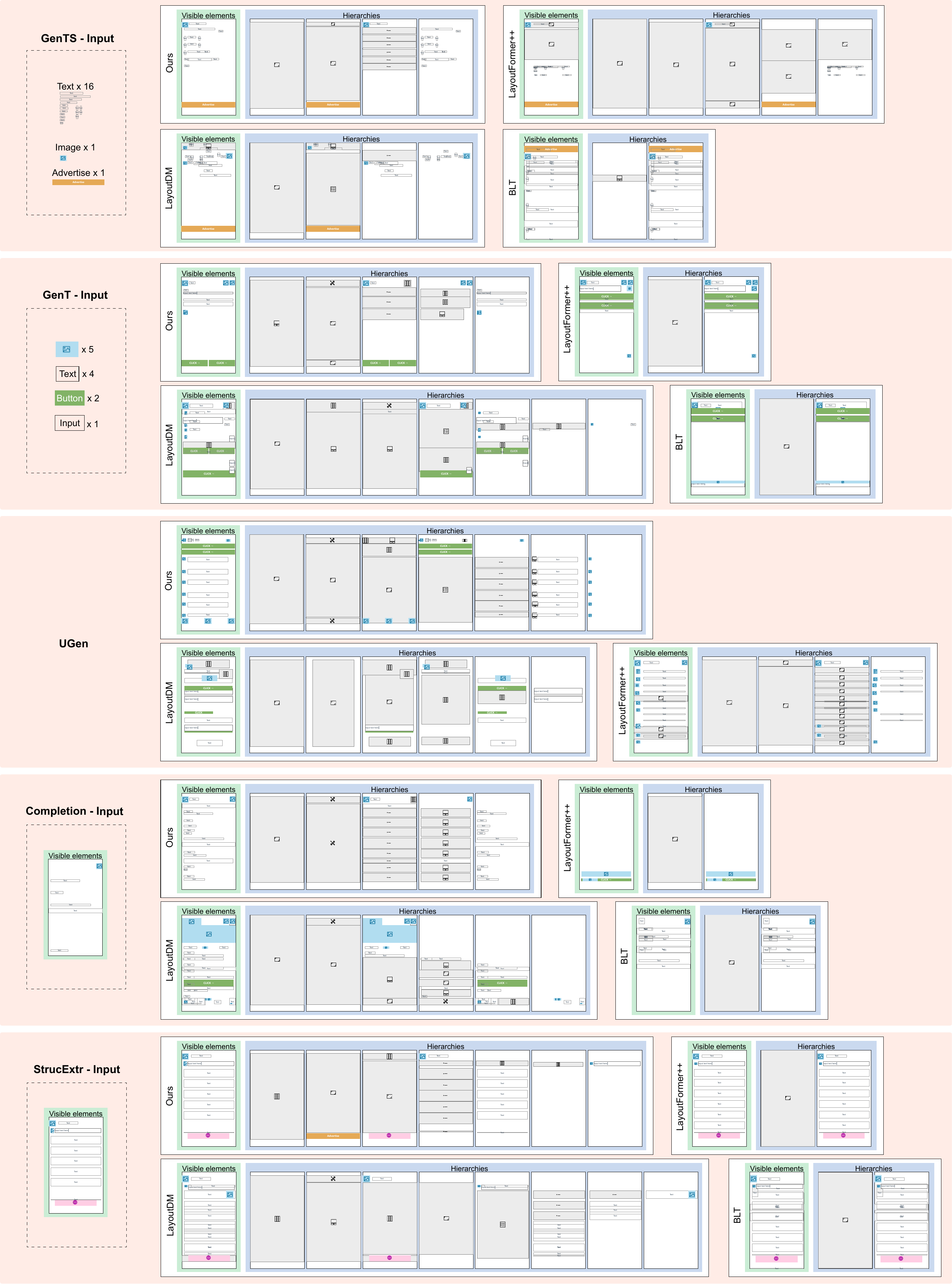}
    \caption{Representative results generated by the compared approaches in the tasks of \GenTS, \GenT, \UGen, \Completion, and \StructExtr\ on RICO. For each sequence of
results, the first one is the visible elements, and the subsequent ones indicate the hierarchies. More results are included in the supplemental material.}
    \label{fig:result-comp-all}
\end{figure*}

\begin{figure*}
    \centering
    \includegraphics[width=\linewidth]{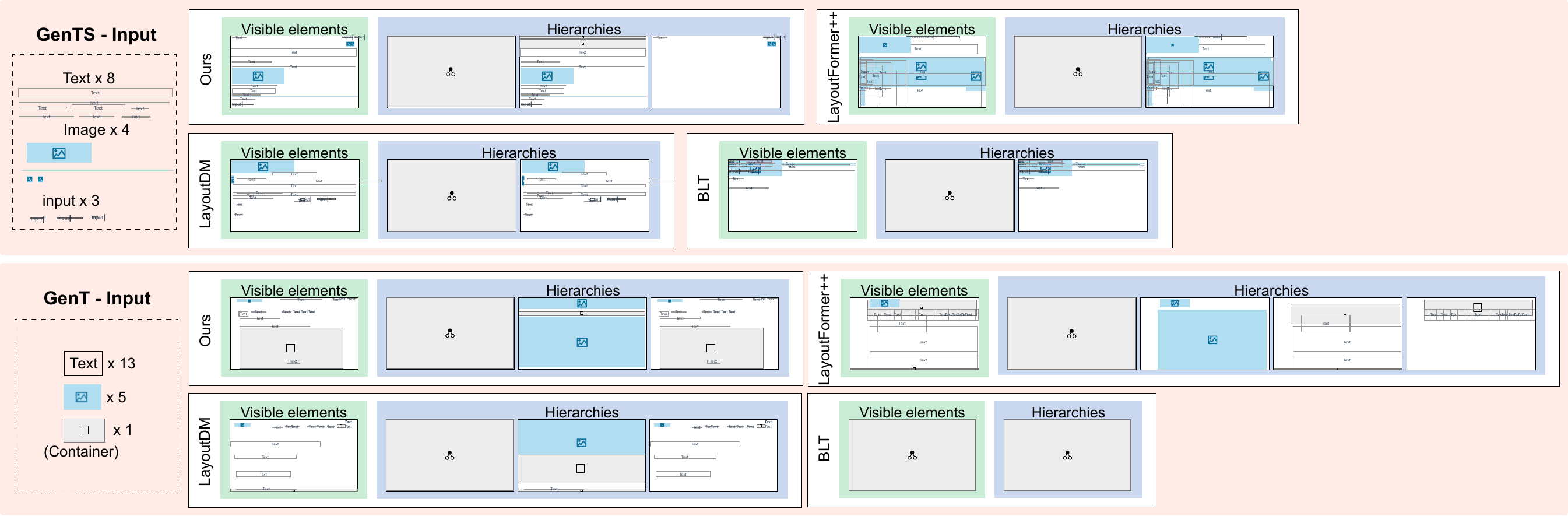}
    \caption{Representative results generated by the compared approaches in the tasks of \GenTS\ and \GenT\ on WebForest.}
    \label{fig:result-comp-web}
\end{figure*}

\begin{figure*}
    \centering
    \includegraphics[width=\linewidth]{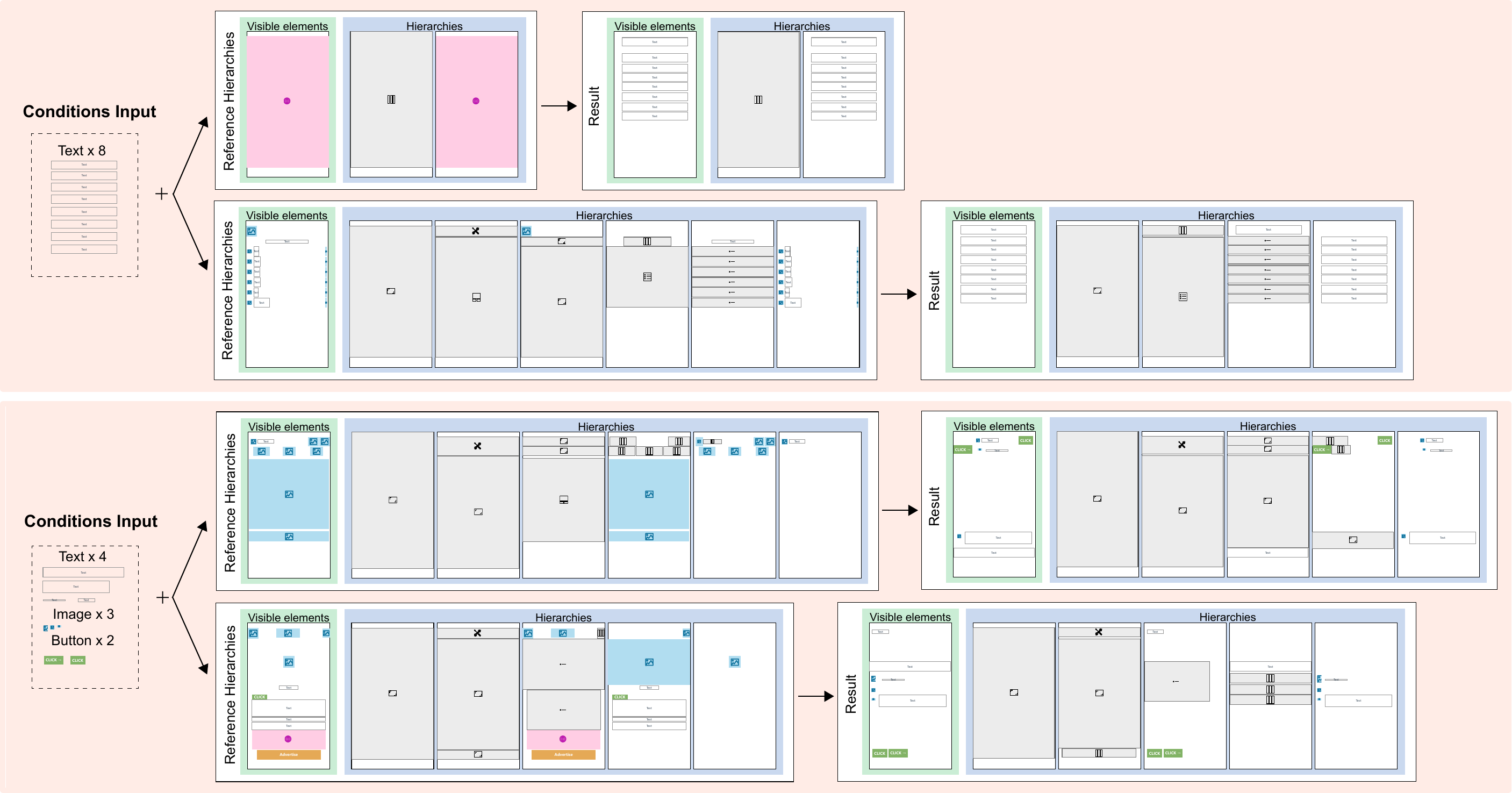}
    \caption{Results of our approach in the task of \StructTran.}
    \label{fig:result-comp-structtran}
\end{figure*}

\begin{figure*}
    \centering
    \includegraphics[width=\linewidth]{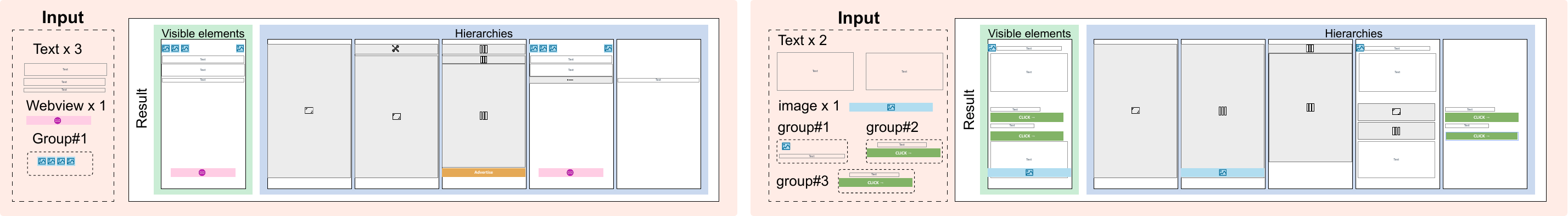}
    \caption{Results of our approach in the task of \GenO.}
    \label{fig:result-comp-geno}
\end{figure*}

\textbf{Evaluation metrics.}
Our approach generates structured layouts represented as layout trees. To evaluate the quality of the generated layouts, we adopt two types of metrics, which we term \EleMetric\ and \StructMetric. The \EleMetric\ measures the quality of element arrangements. We adopt the alignment score (\Align)~\cite{jiang2023layoutformer++}, the overlap score (\Overlap)~\cite{li2020attribute}, the Wasserstein distance for the label distribution (\WLabel)~\cite{arroyo2021variational}, and the Wasserstein distance for the bounding box distribution (\WBox)~\cite{arroyo2021variational}, as the \EleMetric. Since they are about element arrangements, we use the visible graphic elements in layouts for their computations.
The \StructMetric\ measures the quality of layout structures. We find it difficult to measure the quality of global structures. However, we observe that the quality of local structures reflects the quality of global structures. We thus define the following metrics as the \StructMetric. 

\SAlign. 
This metric measures the alignment of local structures. Here, a local structure can be a set of sibling nodes. Given a structured layout, we first collect all its local structures, \ie, all sibling sets. We then compute the alignment score~\cite{jiang2023layoutformer++} of each sibling set and define \SAlign\ of a structured layout as the average of these scores. 

\SOverlap. 
This metric measures the overlap of local structures. A local structure is also defined as a sibling set. Then \SOverlap\ of a structured layout is defined as the average overlap score~\cite{li2020attribute} of its sibling sets. 

\SInclusion.
This metric measures the inclusion of local structures. Here, a local structure is defined as a pair of a parent node and a child node. Since a parent node often serves as the container of its child nodes, the child nodes should be included in the parent node. We thus define the inclusion score of a parent-child pair as the intersection of the parent node and child node over the child node. 
Then, \SInclusion\ of a structured layout is defined as the average inclusion score of all its parent-child pairs. 

\WSLabel\ and \WSBox.
These two metrics reflect whether the generated layout structures are similar to those in the datasets. We use parent-child pairs as local structures to define these two metrics. Specifically, we consider the distribution of label pairs (\WSLabel) and bounding box pairs (\WSBox) to compute the Wasserstein distance~\cite{arroyo2021variational} between real and generated layouts. 

\textbf{Implementation details.}
We implement our approach by PyTorch. The model is trained using the Adam optimizer~\cite{kingmaB2015adam} with NVIDIA RTX 3090 GPUs. The Transformer blocks have 512 embedding dimensions and 2048 feed-forward dimensions. For the conditional layout generator, the Transformer block has 6 layers. For the structure encoder and context encoder, the Transformer blocks have 4 layers.

\subsection{Conditional structured layout generation}

\textbf{Tasks.}
Our approach supports the following existing layout generation tasks: generation conditioned on element types (\GenT); generation conditioned on element types and sizes (\GenTS); completion from {given} elements (\Completion); and unconstrained generation (\UGen). Since our approach considers layout structures explicitly, it also supports the following new tasks: structure extraction from given elements (\StructExtr); layout generation conditioned on element organizations (\GenO); and structure transfer between structured layouts (\StructTran). Below, we briefly describe how our approach achieves these tasks.


\GenT, \GenTS, \Completion, and \UGen. These tasks do not require any structural conditions. When generating a structured layout, the element condition sequence only includes element attribute conditions. The structure code is randomly sampled from the constructed structure space. The randomly sampled structure codes ensure structure diversity in the generated layouts. On the other hand, the structured code can also be predetermined to achieve more controllable structured layout generation.

\StructExtr. This task takes a complete set of elements as conditions. To extract its structure, we randomly sample a structure code for the structured layout generation. To obtain a more reasonable structure, we can estimate a new structure code from the generated structure and use this new code for a new round of structure generation. This procedure may iterate until a better structure is extracted.

\GenO. This task requires element organizations as conditions. When generating a structured layout, the element condition sequence includes element attribute conditions and organization conditions. The structure code is also randomly sampled.

\StructTran. This task transfers the structure of an existing structured layout to an unstructured layout. It is achieved by using the structured layout's structure code and the unstructured layout's elements as conditions.

\textbf{Evaluation settings.}
The existing approaches basically support \GenT, \GenTS, \Completion, and \UGen. In addition, 
we notice that the second step when using the baselines is similar to \StructExtr. 
We thus compare our approach with the baselines on these tasks. Specifically, we compare our approach with (a) LayoutFormer++~\cite{jiang2023layoutformer++}, BLT~\cite{kong2022blt}, and LayoutDM~\cite{inoue2023layoutdm} on \GenT, \GenTS, \Completion, and \StructExtr; (b) LayoutFormer++~\cite{jiang2023layoutformer++}, and LayoutDM~\cite{inoue2023layoutdm} on \UGen. 
In these tasks, we randomly select 1000 layout samples from the testing set and mask specific attributes for each sample to create condition settings. For example, to create the condition settings for \GenT, we mask all the attributes except for the element types for the selected 1000 layout samples.
For each condition setting and each compared approach, we generate a layout. The generated layouts are used for quantitative and qualitative comparisons to show the effectiveness of our approach. Since the baselines can not take structures as input, our approach takes randomly sampled structure codes in these tasks for a fair comparison.

Since no existing approach supports \GenO, we demonstrate qualitative results in this task.
\cite{kikuchi2021modeling} is a computational approach  supporting \StructTran. However, this approach adopts different settings from ours and is not open-source. 

\begin{table*}[t]
    \centering
    \resizebox{\textwidth}{!}{
        
        \begin{tabular}{ l  l  r  r  c  c  c  r  r  r  r }
            \toprule
            & & \multicolumn{5}{ c } {\StructMetricCapital}  & \multicolumn{4}{ c } {\EleMetricCapital} \\
            \cmidrule(lr){3-7}
            \cmidrule(lr){8-11}
            
            Tasks & Approaches & 
            \WSLabel\ $\downarrow$ & \WSBox\ $\downarrow$& \SInclusion\ $\uparrow$& \SAlign\ $\downarrow$& \SOverlap\ $\downarrow$ & \Align\ $\downarrow$& \Overlap\ $\downarrow$& \WLabel\ $\downarrow$& \WBox\ $\downarrow$\\
            \hline
            \multirow{2}{*}{\GenT} & Ours &
                0.69 & \textbf{0.018} & \textbf{0.943} & \textbf{0.0051} & 0.189 & \textbf{0.0015} & 0.065 & 0.345 & \textbf{0.027}
            \\
                
            & GTLayout &
                \textbf{0.56} & 0.057 & 0.938 & 0.0083 & \textbf{0.152} & 0.0018 & \textbf{0.028} & \textbf{0.170} & 0.045 
            \\
            
            \hline

            \multirow{2}{*}{\GenTS} & Ours &
                0.70 & \textbf{0.017} & 0.924 & \textbf{0.0057} & 0.181 & \textbf{0.0011} & 0.066 & 0.348 & \textbf{0.022}
            \\
                
            & GTLayout &
                \textbf{0.62} & 0.071 & \textbf{0.939} & 0.0085 & \textbf{0.146} & 0.0020 & 
                \textbf{0.027} & \textbf{0.110} & 0.049 
            \\
            
            \bottomrule
        \end{tabular}
    }
    \vspace{1mm}
    \caption{Quantitative comparisons between our approach and GTLayout in the tasks of \GenT\ and \GenTS.}
    \label{tab:gtlayComp}
\end{table*}

\begin{figure*}[t]
    \centering
    \includegraphics[width=\linewidth]{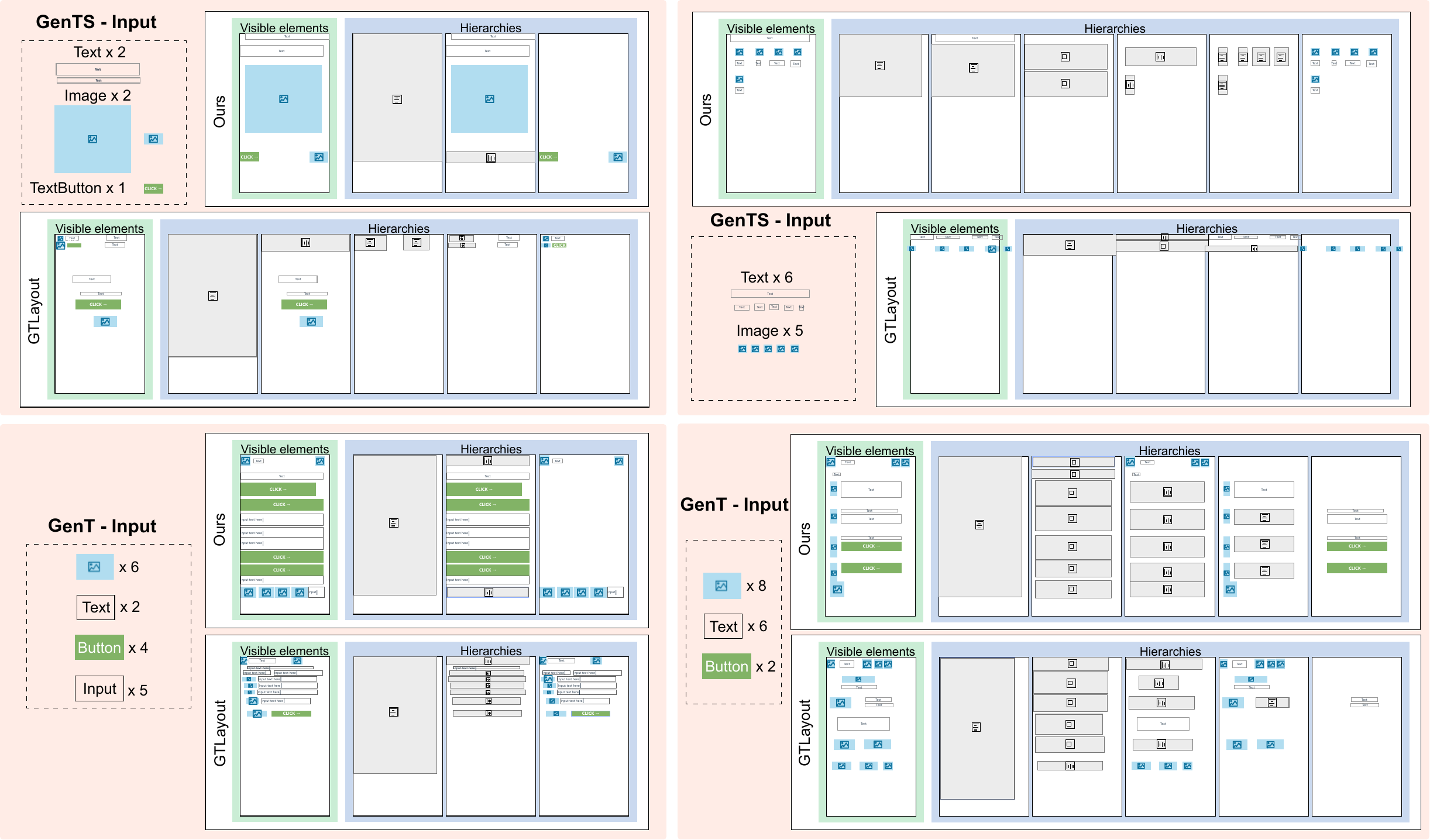}
    \caption{Representative results of our approach and GTLayout in the tasks of \GenT\ and \GenTS.}
    \label{fig:result-comp-gtlay}
\end{figure*}

\textbf{Results and discussion.}
Tables~\ref{tab:rico} and \ref{tab:web} show the quantitative results of the compared approaches in the tasks of \GenT, \GenTS, \UGen, \Completion, and \StructExtr. Our approach achieves the best performance in most \StructMetric. The most representative metrics are \WSLabel\ and \WSBox. These two metrics measure the quality of the generated structures, which is the target of our paper. Our approach achieves the best scores in almost all tasks, confirming the rationality of our generated structures. For \SAlign\ and \SOverlap, our approach achieves good \SOverlap\ scores while keeping \SAlign\ scores comparative to other baselines. this also confirms the high quality of our generated layouts.   
The baselines achieve better scores in \SInclusion. This is reasonable since their layout structures are extracted by maximizing the inclusion scores. This metric would be more meaningful if all the compared approaches could produce structures explicitly. 

The baselines generate structured layouts with three steps and the visible elements are produced in the first step. In contrast, our approach produces layouts' visible elements and structures simultaneously. However, our approach still produces layouts with comparable alignment quality of visible elements. This is confirmed by the \EleMetric.

Figure \ref{fig:result-comp-all} shows the qualitative results of the compared approaches in the task of \GenTS, \GenT, \UGen, \Completion, and \StructExtr\ on RICO. 
For each task, we provide one input setting and the layouts produced by the compared approaches. The results on WebForest have lower qualities since this dataset does not contain sufficient samples. We do not include these results in the paper. More results are included in the supplemental material. These visual results further confirm the superiority of our approach. Our approach can generate samples with high-quality hierarchical structures and visual elements. In contrast, the layout hierarchies generated by LayoutDM, LayoutFormer++, and BLT are less realistic or meaningful. For example, In the task of \StructExtr, our approach generates reasonable hierarchies that appropriately organize the visual elements; LayoutDM produces specious hierarchies that are not functionally justifiable; LayoutFormer++ and BLT even fail to generate any multi-level hierarchies. Figure \ref{fig:result-comp-web} shows the qualitative results of the compared approaches in the task of \GenTS\ and \GenT\ on WebForest. Due to the limited size of the dataset, the results produced by the compared methods are not satisfactory. Nevertheless, our approach still demonstrates superior performance over the baselines in terms of both visible elements and structure.

Figure \ref{fig:result-comp-structtran} and Figure \ref{fig:result-comp-geno} show the results of our approach in the task of \StructTran\ and \GenO. \GenO\ allows people to give high-level organizations of elements. Our model can find appropriate structures for such organizations. This function may help with structured layout design. \StructTran\ helps quickly design structured layouts from existing ones or create a set of structured layouts with similar styles. Figure \ref{fig:result-comp-structtran} demonstrates that the layouts generated by our approach faithfully capture the structure of the reference layouts. These functions further increase the application scenarios of our approach. 

Our approach has the same training and inference time complexity as the other Transformer-based approaches, although our approach includes more attention mask operations. On average, our approach can generate a structured layout in 1.51 seconds. In comparison, to generate a structured layout, LayoutFormer++ costs 1.74 seconds, LayoutDM costs 0.03 seconds, and LayoutTransformer costs 0.11 seconds. As a future work, our approach can be further accelerated by parallelization, which is adopted by LayoutTransformer.

\subsection{Comparison with GTLayout}\label{sec:comp-gtlayout}

GTLayout~\cite{xu2024gtlayout} adopts RvNN for structured layout generation. However, it primarily focuses on layout construction and interpolation without extensively addressing conditional layout generation. We first extend GTLayout to support conditional generation and then compare our approach with GTLayout under various conditional settings.

The GTLayout pipeline consists of a VAE encoder and decoder, which encodes structured layouts into latent space vectors and generates layouts from vectors sampled from this latent space. Inspired by CVAE~\cite{sohn2015learning}, we introduce a conditional encoder before the original VAE decoder. This encoder is a single-layer transformer that integrates the latent code and conditional inputs into a feature vector, which then serves as the input to the VAE decoder. GTLayout uses a dataset that differs from ours and includes distinct internal node types, \ie, vertical arrangement, horizontal arrangement, and stack arrangement. The design of these internal node types is tied to GTLayout's model architecture, making it challenging for the GTLayout pipeline to adapt to our dataset. We thus compare our method with GTLayout using their RICO dataset.

\textbf{Results and discussion.} 
Table~\ref{tab:gtlayComp} shows the quantitative results. These quantitative results demonstrate that our method performs comparatively to GTLayout, confirming the generation ability of both methods. GTLayout even achieves better \Overlap, \SOverlap, \WLabel, and \WSLabel\ scores. Although GTLayout is effective in generating high-quality structured layouts, it often fails to satisfy the input conditions and deviates from the goal of conditional generation.
Figure~\ref{fig:result-comp-gtlay} displays results produced by our method and GTLayout. GTLayout struggles with correctly handling the number and size of input elements, often failing to predict the exact number of nodes and generating unnecessary elements. GTLayout employs relative bounding box representations, which prevent it from accommodating global size constraints. Furthermore, GTLayout is limited to three specific internal nodes: vertical arrangement, horizontal arrangement, and stack arrangement. In contrast, our method can handle datasets with custom internal nodes.

\begin{table*}[t]
    \centering
        \resizebox{\linewidth}{!}{
        \begin{tabular}{  l  c   c  c  c c r  r  r  r }
            \toprule
             & \multicolumn{5}{ c } {\StructMetricCapital}  & \multicolumn{4}{ c } {\EleMetricCapital} \\
            
            \cmidrule(lr){2-6}
            \cmidrule(lr){7-10}
            
            Configurations  & 
            
            \WSLabel\ $\downarrow$ & \WSBox\ $\downarrow$& \SInclusion\ $\uparrow$& \SAlign\ $\downarrow$& \SOverlap\ $\downarrow$ & \Align\ $\downarrow$& \Overlap\ $\downarrow$& \WLabel\ $\downarrow$& \WBox\ $\downarrow$ \\
            \hline
\textbf{Complete model} &
                \textbf{3.33}  & 0.064  & 0.938  & \textbf{0.0017}  & \textbf{0.056} & 0.0037 & \textbf{0.020} & \textbf{0.091} & \textbf{0.045}
            \\

            \textbf{W/o local context} &
                3.96 & 0.063 & \textbf{0.943} & 0.0031 & 0.064 & \textbf{0.0033} & 0.026 & 0.115 & \textbf{0.045}
            \\

            \textbf{W/o global context} &
                6.31 & \textbf{0.032} & 0.923 & 0.0031 & 0.066 & 0.0037 & 0.040 & 0.097 & 0.069
            \\


            
            \bottomrule
        \end{tabular}
    }
    \vspace{1mm}
    \caption{Quantitative results of the ablation study.}
    \label{tab:ablation}
\end{table*}

\begin{figure*}[t]
    \centering
    \includegraphics[width=\linewidth]{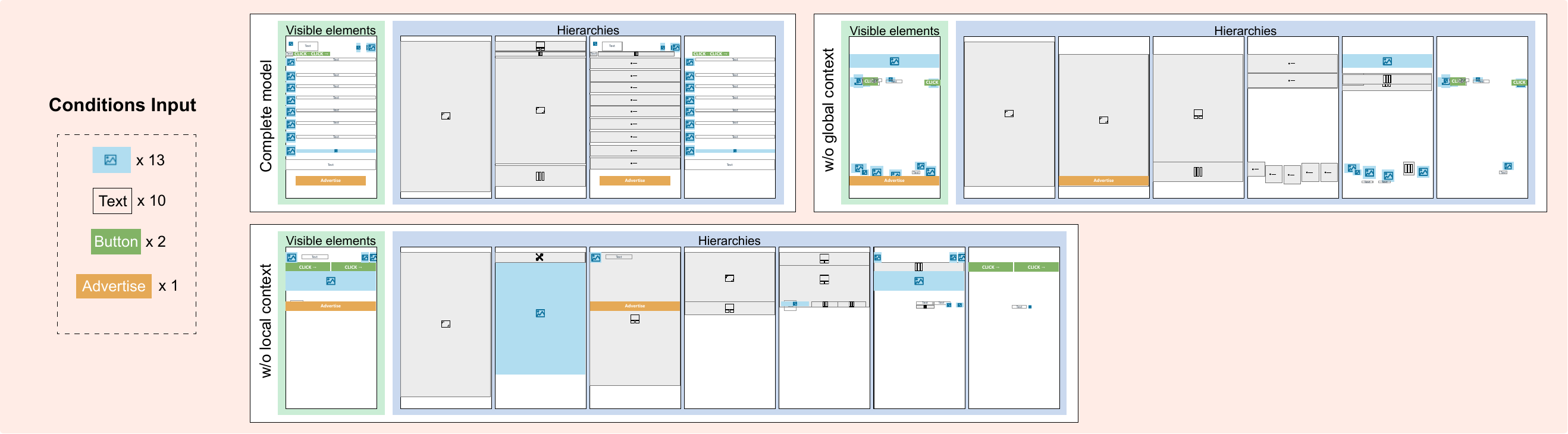}
    \caption{Representative results of the ablation study.}
    \label{fig:result-ablation}
\end{figure*}

\subsection{Ablation study}



We conduct an ablation study to demonstrate the necessity of the local context and the global context. In this study, we remove the local context and the global context from the complete model, respectively, and adopt the \GenT\ task for the evaluation. Table~\ref{tab:ablation} shows the quantitative results. The results confirm that the complete model outperforms the other configurations. Without the global context, the model achieves a better \WSBox\ score. It is reasonable since the model predicts the current node based on the parent and the sibling, which is learned from the dataset's distribution of parent-child pairs. Without the local context, the model achieves a better \Align\ score since this score is computed based on global alignment, which may be affected by the local context. Figure~\ref{fig:result-ablation} shows some qualitative results. These results further confirm that the complete model produces layouts with higher quality. Without the local context, the produced layout has defects in local element arrangement; without the global context, the global arrangement of the elements in the produced layout is severely affected. 

We also conduct an experiment on the conditional layout generator. In this experiment, we remove the attention mask, and the model fails to produce reasonable results. For the other component of the conditional layout generator, \eg, the structure code $z$ and the organization token $\langle \mathrm{oc} \rangle$, can not be removed since they are explicitly designed for the specific functions of our approach. For example, without the structure code, our approach can not achieve the \StructTran\ task; without the organization token, our approach can not support organization conditions.

\subsection{Structure for layout optimization}

In this section, we demonstrate that the structured layouts generated by our approach can be used for layout optimization, which is crucial for adapting graphic layouts to different display configurations. 

The structured layouts generated by our approach contain detailed hierarchical organizations of visible graphic elements. Such hierarchical organizations can not be used for layout geometry optimization directly. However, they can help determine the alignment relations among the elements. In a structured layout, an internal node specifies the arrangement types of its children nodes. For example, if an internal node has the type of linear arrangement, then its children elements should be arranged in a sequence. The arrangement direction and order can be determined by the initial geometry of the children elements. With this information, we can extract the precise alignment relations among the elements and use the existing approaches~\cite{xu2019global} for layout optimization.

\begin{figure*}
    \centering
    \includegraphics[width=\linewidth]{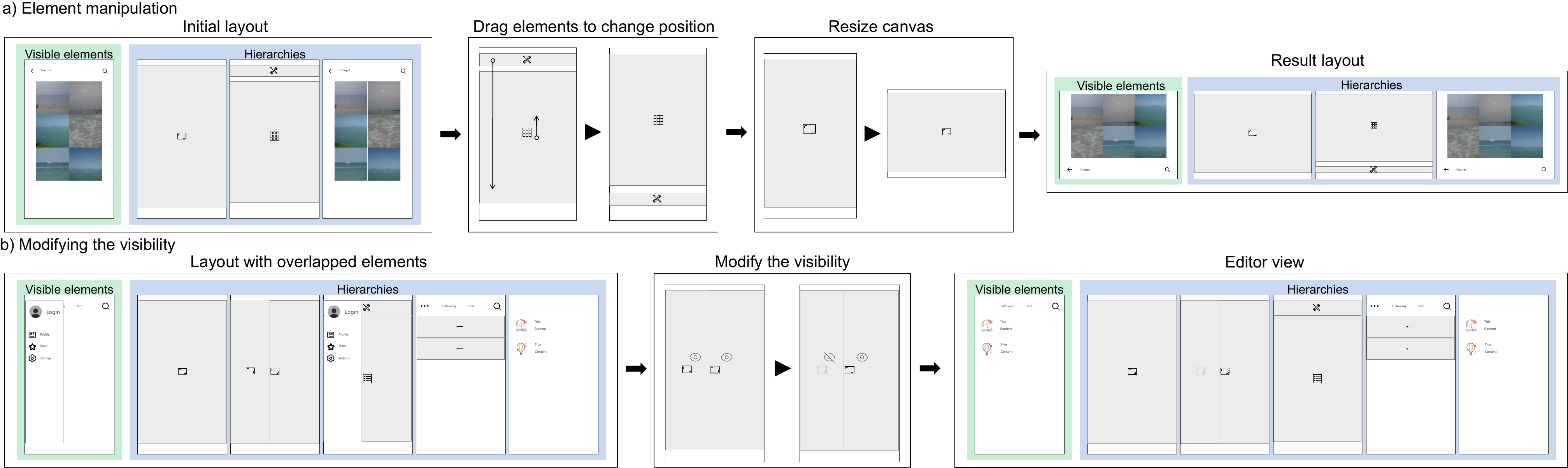}
    \caption{
    The structural information enables easy manipulation of graphic layouts. a) Dragging an element can easily modify its relative position in the layout; resizing an element helps automatically re-arrange its children elements, adapting to different display configurations. b) Modifying the visibility of an internal element eases the manipulation of the layout containing overlapped elements.
    }
    
    \label{fig:userediting}
    
\end{figure*}

The structural information also enables easy manipulation of graphic layouts, which is important in graphic layout design. 
Figure~\ref{fig:userediting} demonstrates layout manipulation by editing visible and internal elements. For example, dragging a visible element can easily modify its relative position in the layout; resizing an internal element helps automatically re-arrange its children elements, adapting to different display configurations; modifying the visibility of an internal element eases the manipulation of the layout containing overlapped elements.

\section{Conclusion}

In this paper, we have presented \techName, a novel Transformer-based approach for conditional structured layout generation. We use a structure serialization scheme to represent structured layouts as sequences. We also disentangle the structural information of layouts from element arrangements, thus achieving better control of layout structures. 
The experiments have confirmed that our approach is more effective in generating structured layouts with conditions compared with the baselines. We have also demonstrated that our approach can achieve layout structure extraction and transfer and discussed the potential applications. To the best of our knowledge, our approach is the first data-driven approach that achieves conditional structured layout generation.


Our approach still has limitations. Given a specific set of conditions for layout generation, the randomly sampled structure code may not always be suitable. For example, with a structure code corresponding to a simple structure and a set of conditions specifying a large number of elements, the generated layout may contain an unexpected element overlay. Although we propose an iterative generation strategy for this problem, it is still necessary to devise a mapping between the conditions and the structures.

In future work, we plan to extend the model for other structured data generation tasks, \eg, 3D shapes and indoor scenes. Our approach requires layouts to be completely structured. It would be meaningful to learn structural patterns from partially structured layouts. It would also be interesting to exploit Diffusion models for structured layout generation.

\section*{Acknowledgements}

We thank the reviewers for their insightful comments. This work was partially supported by grants from NSFC (62472287, 62072316, U21B2023), Guangdong Basic and Applied Basic Research Foundation (2023A1515011297, 2023B1515120026), DEGP Innovation Team (2022KCXTD025), and Scientific Development Funds from Shenzhen University.

{
\bibliographystyle{IEEEtran}
\bibliography{main}

\begin{thebibliography}{10}
\providecommand{\url}[1]{#1}
\csname url@samestyle\endcsname
\providecommand{\newblock}{\relax}
\providecommand{\bibinfo}[2]{#2}
\providecommand{\BIBentrySTDinterwordspacing}{\spaceskip=0pt\relax}
\providecommand{\BIBentryALTinterwordstretchfactor}{4}
\providecommand{\BIBentryALTinterwordspacing}{\spaceskip=\fontdimen2\font plus
\BIBentryALTinterwordstretchfactor\fontdimen3\font minus \fontdimen4\font\relax}
\providecommand{\BIBforeignlanguage}[2]{{%
\expandafter\ifx\csname l@#1\endcsname\relax
\typeout{** WARNING: IEEEtran.bst: No hyphenation pattern has been}%
\typeout{** loaded for the language `#1'. Using the pattern for}%
\typeout{** the default language instead.}%
\else
\language=\csname l@#1\endcsname
\fi
#2}}
\providecommand{\BIBdecl}{\relax}
\BIBdecl

\bibitem{xu2019global}
P.~Xu, G.~Yan, H.~Fu, T.~Igarashi, C.-L. Tai, and H.~Huang, ``Global beautification of 2d and 3d layouts with interactive ambiguity resolution,'' \emph{IEEE transactions on visualization and computer graphics}, vol.~27, no.~4, pp. 2355--2368, 2019.

\bibitem{zeidler2013auckland}
C.~Zeidler, C.~Lutteroth, W.~Sturzlinger, and G.~Weber, ``The auckland layout editor: An improved gui layout specification process,'' in \emph{Proceedings of the 26th annual ACM symposium on User interface software and technology}, 2013, pp. 343--352.

\bibitem{dayama2020grids}
N.~R. Dayama, K.~Todi, T.~Saarelainen, and A.~Oulasvirta, ``Grids: Interactive layout design with integer programming,'' in \emph{Proceedings of the 2020 CHI Conference on Human Factors in Computing Systems}, 2020, pp. 1--13.

\bibitem{swearngin2020scout}
A.~Swearngin, C.~Wang, A.~Oleson, J.~Fogarty, and A.~J. Ko, ``Scout: Rapid exploration of interface layout alternatives through high-level design constraints,'' in \emph{Proceedings of the 2020 CHI Conference on Human Factors in Computing Systems}, 2020, pp. 1--13.

\bibitem{xu2022hierarchical}
P.~Xu, Y.~Li, Z.~Yang, W.~Shi, H.~Fu, and H.~Huang, ``Hierarchical layout blending with recursive optimal correspondence,'' \emph{ACM Transactions on Graphics (Proceedings of SIGGRAPH ASIA)}, vol.~41, no.~6, pp. 249:1--249:15, 2022.

\bibitem{arroyo2021variational}
D.~M. Arroyo, J.~Postels, and F.~Tombari, ``Variational transformer networks for layout generation,'' in \emph{Proceedings of the IEEE/CVF Conference on Computer Vision and Pattern Recognition}, 2021, pp. 13\,642--13\,652.

\bibitem{chai2023layoutdm}
S.~Chai, L.~Zhuang, and F.~Yan, ``Layoutdm: Transformer-based diffusion model for layout generation,'' in \emph{Proceedings of the IEEE/CVF Conference on Computer Vision and Pattern Recognition}, 2023, pp. 18\,349--18\,358.

\bibitem{gupta2021layouttransformer}
K.~Gupta, J.~Lazarow, A.~Achille, L.~S. Davis, V.~Mahadevan, and A.~Shrivastava, ``Layouttransformer: Layout generation and completion with self-attention,'' in \emph{Proceedings of the IEEE/CVF International Conference on Computer Vision}, 2021, pp. 1004--1014.

\bibitem{hui2023unifying}
M.~Hui, Z.~Zhang, X.~Zhang, W.~Xie, Y.~Wang, and Y.~Lu, ``Unifying layout generation with a decoupled diffusion model,'' in \emph{Proceedings of the IEEE/CVF Conference on Computer Vision and Pattern Recognition}, 2023, pp. 1942--1951.

\bibitem{inoue2023layoutdm}
N.~Inoue, K.~Kikuchi, E.~Simo-Serra, M.~Otani, and K.~Yamaguchi, ``{LayoutDM: Discrete Diffusion Model for Controllable Layout Generation},'' in \emph{Proceedings of the IEEE/CVF Conference on Computer Vision and Pattern Recognition}, 2023, pp. 10\,167--10\,176.

\bibitem{jiang2022coarse}
Z.~Jiang, S.~Sun, J.~Zhu, J.-G. Lou, and D.~Zhang, ``Coarse-to-fine generative modeling for graphic layouts,'' in \emph{AAAI'22}, February 2022.

\bibitem{jyothi2019layoutvae}
A.~A. Jyothi, T.~Durand, J.~He, L.~Sigal, and G.~Mori, ``Layoutvae: Stochastic scene layout generation from a label set,'' in \emph{Proceedings of the IEEE/CVF International Conference on Computer Vision}, 2019, pp. 9895--9904.

\bibitem{kikuchi2021constrained}
K.~Kikuchi, E.~Simo-Serra, M.~Otani, and K.~Yamaguchi, ``Constrained graphic layout generation via latent optimization,'' in \emph{Proceedings of the 29th ACM International Conference on Multimedia}, 2021, pp. 88--96.

\bibitem{kong2022blt}
X.~Kong, L.~Jiang, H.~Chang, H.~Zhang, Y.~Hao, H.~Gong, and I.~Essa, ``Blt: bidirectional layout transformer for controllable layout generation,'' in \emph{Computer Vision--ECCV 2022: 17th European Conference, Tel Aviv, Israel, October 23--27, 2022, Proceedings, Part XVII}.\hskip 1em plus 0.5em minus 0.4em\relax Springer, 2022, pp. 474--490.

\bibitem{zheng2019content}
X.~Zheng, X.~Qiao, Y.~Cao, and R.~W. Lau, ``Content-aware generative modeling of graphic design layouts,'' \emph{ACM Transactions on Graphics (TOG)}, vol.~38, no.~4, pp. 1--15, 2019.

\bibitem{goodfellow2014generative}
\BIBentryALTinterwordspacing
I.~Goodfellow, J.~Pouget-Abadie, M.~Mirza, B.~Xu, D.~Warde-Farley, S.~Ozair, A.~Courville, and Y.~Bengio, ``Generative adversarial nets,'' in \emph{Advances in Neural Information Processing Systems}, Z.~Ghahramani, M.~Welling, C.~Cortes, N.~Lawrence, and K.~Weinberger, Eds., vol.~27.\hskip 1em plus 0.5em minus 0.4em\relax Curran Associates, Inc., 2014. [Online]. Available: \url{https://proceedings.neurips.cc/paper/2014/file/5ca3e9b122f61f8f06494c97b1afccf3-Paper.pdf}
\BIBentrySTDinterwordspacing

\bibitem{kingma2013auto}
D.~P. Kingma and M.~Welling, ``Auto-encoding variational bayes,'' in \emph{International Conference on Learning Representations}, 2013.

\bibitem{scarselli2008graph}
F.~Scarselli, M.~Gori, A.~C. Tsoi, M.~Hagenbuchner, and G.~Monfardini, ``The graph neural network model,'' \emph{IEEE Transactions on Neural Networks}, vol.~20, no.~1, pp. 61--80, 2008.

\bibitem{vaswani2017attention}
\BIBentryALTinterwordspacing
A.~Vaswani, N.~Shazeer, N.~Parmar, J.~Uszkoreit, L.~Jones, A.~N. Gomez, L.~u. Kaiser, and I.~Polosukhin, ``Attention is all you need,'' in \emph{Advances in Neural Information Processing Systems}, I.~Guyon, U.~V. Luxburg, S.~Bengio, H.~Wallach, R.~Fergus, S.~Vishwanathan, and R.~Garnett, Eds., vol.~30.\hskip 1em plus 0.5em minus 0.4em\relax Curran Associates, Inc., 2017. [Online]. Available: \url{https://proceedings.neurips.cc/paper/2017/file/3f5ee243547dee91fbd053c1c4a845aa-Paper.pdf}
\BIBentrySTDinterwordspacing

\bibitem{ho2020denoising}
J.~Ho, A.~Jain, and P.~Abbeel, ``Denoising diffusion probabilistic models,'' \emph{Advances in Neural Information Processing Systems}, vol.~33, pp. 6840--6851, 2020.

\bibitem{jiang2019orc}
Y.~Jiang, R.~Du, C.~Lutteroth, and W.~Stuerzlinger, ``Orc layout: Adaptive gui layout with or-constraints,'' in \emph{Proceedings of the 2019 CHI Conference on Human Factors in Computing Systems}, 2019, pp. 1--12.

\bibitem{deka2017rico}
B.~Deka, Z.~Huang, C.~Franzen, J.~Hibschman, D.~Afergan, Y.~Li, J.~Nichols, and R.~Kumar, ``Rico: A mobile app dataset for building data-driven design applications,'' in \emph{Proceedings of the 30th Annual ACM Symposium on User Interface Software and Technology}, 2017, pp. 845--854.

\bibitem{kikuchi2021modeling}
K.~Kikuchi, M.~Otani, K.~Yamaguchi, and E.~Simo-Serra, ``Modeling visual containment for web page layout optimization,'' in \emph{Computer Graphics Forum}, vol.~40, no.~7.\hskip 1em plus 0.5em minus 0.4em\relax Wiley Online Library, 2021, pp. 33--44.

\bibitem{socher2011parsing}
R.~Socher, C.~C. Lin, C.~Manning, and A.~Y. Ng, ``Parsing natural scenes and natural language with recursive neural networks,'' in \emph{Proceedings of the 28th international conference on machine learning (ICML-11)}, 2011, pp. 129--136.

\bibitem{li2017grass}
J.~Li, K.~Xu, S.~Chaudhuri, E.~Yumer, H.~Zhang, and L.~Guibas, ``Grass: Generative recursive autoencoders for shape structures,'' \emph{ACM Transactions on Graphics (TOG)}, vol.~36, no.~4, pp. 1--14, 2017.

\bibitem{mo2019structurenet}
K.~Mo, P.~Guerrero, L.~Yi, H.~Su, P.~Wonka, N.~Mitra, and L.~J. Guibas, ``Structurenet: Hierarchical graph networks for 3d shape generation,'' \emph{arXiv preprint arXiv:1908.00575}, 2019.

\bibitem{zhu2018scores}
C.~Zhu, K.~Xu, S.~Chaudhuri, R.~Yi, and H.~Zhang, ``Scores: Shape composition with recursive substructure priors,'' \emph{ACM Transactions on Graphics (TOG)}, vol.~37, no.~6, pp. 1--14, 2018.

\bibitem{li2019grains}
M.~Li, A.~G. Patil, K.~Xu, S.~Chaudhuri, O.~Khan, A.~Shamir, C.~Tu, B.~Chen, D.~Cohen-Or, and H.~Zhang, ``Grains: Generative recursive autoencoders for indoor scenes,'' \emph{ACM Transactions on Graphics (TOG)}, vol.~38, no.~2, pp. 1--16, 2019.

\bibitem{jiang2023layoutformer++}
Z.~Jiang, J.~Guo, S.~Sun, H.~Deng, Z.~Wu, V.~Mijovic, Z.~J. Yang, J.-G. Lou, and D.~Zhang, ``Layoutformer++: Conditional graphic layout generation via constraint serialization and decoding space restriction,'' in \emph{Proceedings of the IEEE/CVF Conference on Computer Vision and Pattern Recognition}, 2023, pp. 18\,403--18\,412.

\bibitem{dixon2011content}
M.~Dixon, D.~Leventhal, and J.~Fogarty, ``Content and hierarchy in pixel-based methods for reverse engineering interface structure,'' in \emph{Proceedings of the SIGCHI Conference on Human Factors in Computing Systems}, 2011, pp. 969--978.

\bibitem{jiang2021reverseorc}
Y.~Jiang, W.~Stuerzlinger, and C.~Lutteroth, ``Reverseorc: Reverse engineering of resizable user interface layouts with or-constraints,'' in \emph{Proceedings of the 2021 CHI Conference on Human Factors in Computing Systems}, 2021, pp. 1--18.

\bibitem{o2014learning}
P.~O’Donovan, A.~Agarwala, and A.~Hertzmann, ``Learning layouts for single-page graphic designs,'' \emph{IEEE transactions on visualization and computer graphics}, vol.~20, no.~8, pp. 1200--1213, 2014.

\bibitem{xu2024gtlayout}
P.~Xu, W.~Shi, X.~Hu, H.~Fu, and H.~Huang, ``Gtlayout: Learning general trees for structured grid layout generation,'' in \emph{International Conference on Computational Visual Media}.\hskip 1em plus 0.5em minus 0.4em\relax Springer, 2024, pp. 131--153.

\bibitem{li2019layoutgan}
J.~Li, J.~Yang, A.~Hertzmann, J.~Zhang, and T.~Xu, ``Layoutgan: Generating graphic layouts with wireframe discriminators,'' \emph{arXiv preprint arXiv:1901.06767}, 2019.

\bibitem{lee2020neural}
H.-Y. Lee, L.~Jiang, I.~Essa, P.~B. Le, H.~Gong, M.-H. Yang, and W.~Yang, ``Neural design network: Graphic layout generation with constraints,'' in \emph{European Conference on Computer Vision}.\hskip 1em plus 0.5em minus 0.4em\relax Springer, 2020, pp. 491--506.

\bibitem{kenton2019bert}
J.~D. M.-W.~C. Kenton and L.~K. Toutanova, ``Bert: Pre-training of deep bidirectional transformers for language understanding,'' in \emph{Proceedings of NAACL-HLT}, 2019, pp. 4171--4186.

\bibitem{patil2020read}
A.~G. Patil, O.~Ben-Eliezer, O.~Perel, and H.~Averbuch-Elor, ``Read: Recursive autoencoders for document layout generation,'' in \emph{Proceedings of the IEEE/CVF Conference on Computer Vision and Pattern Recognition Workshops}, 2020, pp. 544--545.

\bibitem{feng2023layoutgpt}
W.~Feng, W.~Zhu, T.-j. Fu, V.~Jampani, A.~Akula, X.~He, S.~Basu, X.~E. Wang, and W.~Y. Wang, ``Layoutgpt: Compositional visual planning and generation with large language models,'' \emph{Advances in Neural Information Processing Systems}, vol.~36, pp. 18\,225--18\,250, 2023.

\bibitem{tang2023layoutnuwa}
Z.~Tang, C.~Wu, J.~Li, and N.~Duan, ``Layoutnuwa: Revealing the hidden layout expertise of large language models,'' \emph{arXiv preprint arXiv:2309.09506}, 2023.

\bibitem{yang2024posterllava}
T.~Yang, Y.~Luo, Z.~Qi, Y.~Wu, Y.~Shan, and C.~W. Chen, ``Posterllava: Constructing a unified multi-modal layout generator with llm,'' \emph{arXiv preprint arXiv:2406.02884}, 2024.

\bibitem{yang2024llplace}
Y.~Yang, J.~Lu, Z.~Zhao, Z.~Luo, J.~J. Yu, V.~Sanchez, and F.~Zheng, ``Llplace: The 3d indoor scene layout generation and editing via large language model,'' \emph{arXiv preprint arXiv:2406.03866}, 2024.

\bibitem{lin2023parse}
J.~Lin, J.~Guo, S.~Sun, W.~Xu, T.~Liu, J.-G. Lou, and D.~Zhang, ``A parse-then-place approach for generating graphic layouts from textual descriptions,'' in \emph{Proceedings of the IEEE/CVF International Conference on Computer Vision}, 2023, pp. 23\,622--23\,631.

\bibitem{laurenccon2024unlocking}
H.~Lauren{\c{c}}on, L.~Tronchon, and V.~Sanh, ``Unlocking the conversion of web screenshots into html code with the websight dataset,'' \emph{arXiv preprint arXiv:2403.09029}, 2024.

\bibitem{li2020auto}
Y.~Li, J.~Amelot, X.~Zhou, S.~Bengio, and S.~Si, ``Auto completion of user interface layout design using transformer-based tree decoders,'' \emph{arXiv preprint arXiv:2001.05308}, 2020.

\bibitem{li2020attribute}
J.~Li, J.~Yang, J.~Zhang, C.~Liu, C.~Wang, and T.~Xu, ``Attribute-conditioned layout gan for automatic graphic design,'' \emph{IEEE Transactions on Visualization and Computer Graphics}, vol.~27, no.~10, pp. 4039--4048, 2020.

\bibitem{kingmaB2015adam}
\BIBentryALTinterwordspacing
D.~P. Kingma and J.~Ba, ``Adam: {A} method for stochastic optimization,'' in \emph{3rd International Conference on Learning Representations, {ICLR} 2015, San Diego, CA, USA, May 7-9, 2015, Conference Track Proceedings}, Y.~Bengio and Y.~LeCun, Eds., 2015. [Online]. Available: \url{http://arxiv.org/abs/1412.6980}
\BIBentrySTDinterwordspacing

\bibitem{sohn2015learning}
K.~Sohn, H.~Lee, and X.~Yan, ``Learning structured output representation using deep conditional generative models,'' \emph{Advances in neural information processing systems}, vol.~28, 2015.

\end{thebibliography}
}
\vspace{-20pt}

\begin{IEEEbiography}[{\includegraphics[width=1in,height=1.25in,clip,keepaspectratio]{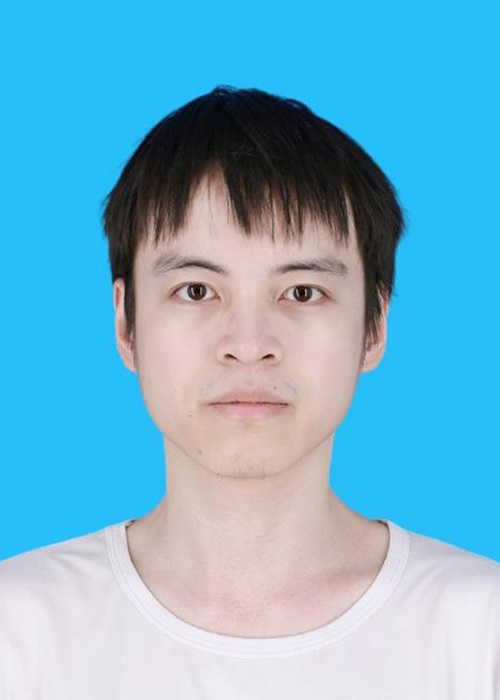}}]{Xin Hu} is an M.Sc. candidate in the Visual Computing Research Center at Shenzhen University, China. He received his bachelor's degree in computer science from Jilin University in 2022. His research interest is computer graphics.
\end{IEEEbiography}

\vspace{-20pt}

\begin{IEEEbiography}[{\includegraphics[width=1in,height=1.25in,clip,keepaspectratio]{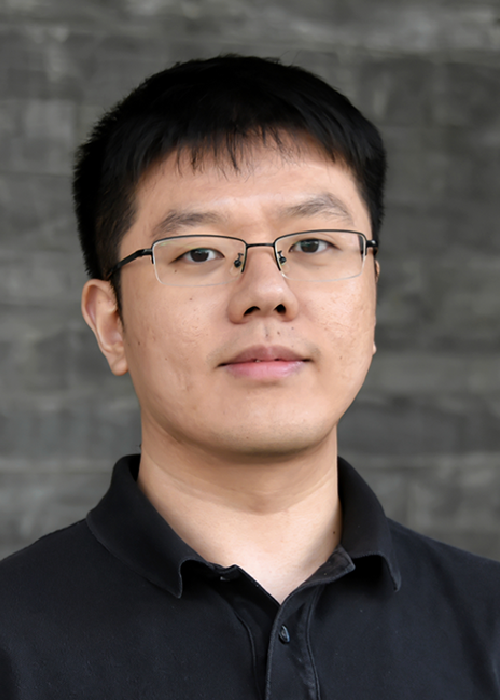}}]{Pengfei Xu} is an Associate Professor of the College of Computer Science and Software Engineering at Shenzhen University. He received his Bachelor's degree in Math from Zhejiang University, China, in 2009 and his Ph.D. in Computer Science from the Hong Kong University of Science and Technology in 2015. His primary research lies in Human-Computer Interaction and Computer Graphics.
\end{IEEEbiography}

\vspace{-20pt}

\begin{IEEEbiography}[{\includegraphics[width=1in,height=1.25in,clip,keepaspectratio]{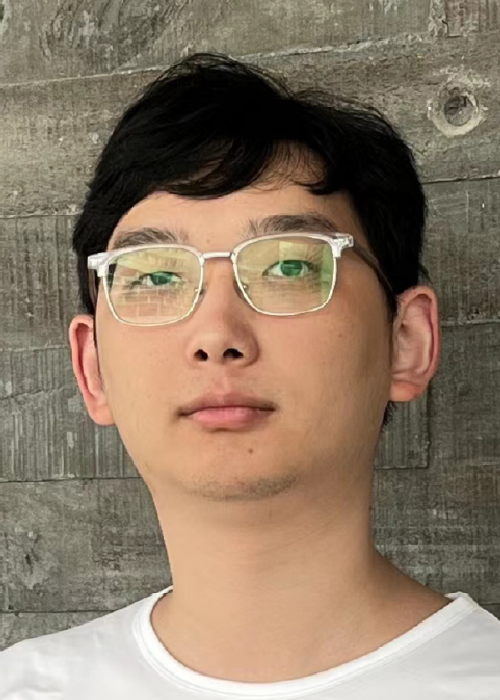}}]{Jin Zhou} is a Master’s candidate at the Visual Computing Center, Shenzhen University, specializing in computer graphics. He earned his Bachelor’s degree from Shanghai Maritime University in 2023.  His research interest is computer graphics.
\end{IEEEbiography}

\vspace{-20pt}

\begin{IEEEbiography}[{\includegraphics[width=1in,height=1.25in,clip,keepaspectratio]{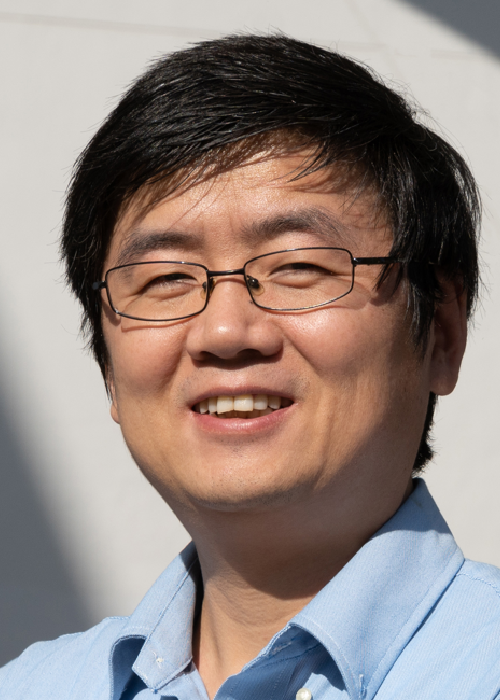}}]{Hongbo Fu}
received a BS degree in information sciences from Peking University, China, in 2002 and a PhD degree in computer science from the Hong Kong University of Science and Technology in 2007. He is a Full Professor of the Division of Emerging Interdisciplinary Areas at the Hong Kong University of Science and Technology. His primary research interests fall in the fields of computer graphics and human-computer interaction. He has served as an Associate Editor of The Visual Computer, Computers \& Graphics, and Computer Graphics Forum.
\end{IEEEbiography}

\vspace{-20pt}

\begin{IEEEbiography}[{\includegraphics[width=1in,height=1.25in,clip,keepaspectratio]{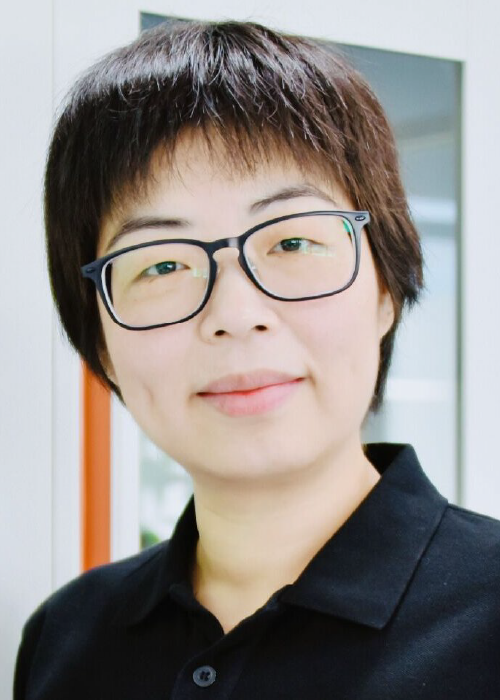}}]{Hui Huang} is a Distinguished TFA Professor at Shenzhen University, where she directs the Visual Computing Research Center. She received her Ph.D. degree in applied math from The University of British Columbia in 2008. Her research interests span computer graphics, vision, and visualization. She is currently a Senior Member of IEEE/ACM/CSIG, a Distinguished Member of CCF, and is on the editorial boards of ACM TOG and IEEE TVCG.
\end{IEEEbiography}



\end{document}